\documentclass[journal=jacsat,manuscript=article]{achemso}
\usepackage{chemformula} 
\usepackage[T1]{fontenc} 
\usepackage[version=3]{mhchem} 

\usepackage{ulem}

\usepackage[colorlinks=true, allcolors=blue]{hyperref}
\usepackage{graphicx}
\usepackage{siunitx}
\usepackage{titlesec} 
\usepackage{amsmath}
\usepackage{cancel}
\usepackage{caption}
\usepackage{multirow}
\usepackage{array}
\usepackage{subcaption}
\usepackage{csquotes}
\usepackage{parskip}
\usepackage{dcolumn}
\usepackage{bm}

\author{Sarga P K}
\author{Karthik H J}
\author{Swastibrata Bhattacharyya}
 \affiliation{Department of Physics, Birla Institute of Technology and Science Pilani, K. K. Birla Goa Campus, Zuarinagar, Goa, 403726, India}
\email{swastibratab@goa.bits-pilani.ac.in}

\title[\textsf{achemso}]{Electronic and Interfacial Properties of 2D MXene/blue phosphorene Heterostructures: Impact of External Strain for Thermoelectric Applications}

\abbreviations{IR,NMR,UV}

\begin{document}

\begin{abstract}

Building two-dimensional (2D) van der Waals (vdW) heterostructures and enhancing their properties through strain engineering unlocks new applications for their constituent materials. In this study, we present a comprehensive first-principles investigation of oxygen-functionalized MXene-based heterostructures (M$_2$CO$_2$ (M=Sc,Zr,Hf)/blue phosphorene), emphasizing their structural, electronic, and thermoelectric properties under the application of various types of strain. Our results indicate a reduction in band gap under strain and metallic transition for in-plane strain (uni- and biaxial strain). Transition from type-I to type-II could be obtained for Hf$_2$CO$_2$/blueP  and Zr$_2$CO$_2$/blueP heterostructure by applying strain, providing a method for their potential application in photocatalytic devices that require type-II band alignment for photogenerated charge separation. We observe a positive correlation between strain and thermoelectric efficiency under most conditions, with a significant enhancement in thermoelectric power factor (PF) and electronic figure of merit (ZT$_e$) achievable in all heterostructures through strain engineering. Among the three heterostructures, Sc$_2$CO$_2$/blueP showed the maximum PF (ZT$_e$) of 11.2 x10$^{11}$ W/mK$^2$s (43.2) for 2\% normal compressive (4\% biaxial compressive) strain. These findings suggest that MXene/blueP heterostructures hold significant promise for applications in optoelectronics and high-temperature thermoelectric devices.

\vspace{5mm}\textbf{keywords} : 2D materials, vdW heterostructure, Structural and electronic properties, Strain engineering, Modulation of electronic properties, Thermoelectric application, First-principles calculations.
\end{abstract}

\section{INTRODUCTION}

After the successful extraction of graphene from graphite using a mechanical exfoliation method\cite{geim2007rise}, research into two-dimensional (2D) materials with unique electrical and optical properties has flourished.  When it comes to materials science, chemistry, nanoscience, optics, and other subjects, 
2D materials shine\cite{liu2020two,pomerantseva2017two} due to their unique properties driven by their dimension compared to their bulk form. Different 2D materials, including  graphene, transition metal dichalcogenides (TMDCs), MXene, hexagonal BN (h-BN), and phosphorene are attracting greater attention.
MXenes are a new class of 2D transition-metal carbides/carbonitrides/nitrides/borides. Due to their wide range of appealing features, these materials have exhibited considerable potential in many applications such as batteries, capacitors, and electrocatalysis.\cite{naguib2012two,naguib201425th,anasori2015two} MXenes are derived from MAX phases through etching of the weakly bonded "A" layer. MAX phases are thermodynamically stable, layered hexagonal materials with the general formula M$_{n+1}$AX$_n$ (n=1,2, or 3), where \enquote{M} is an early transition metal (Ti, Sc, Hf, Zr, V, Nb, Ta, Cr, or Mo), \enquote{A} is an A-group element (mostly groups 13 and 14), and \enquote{X} is carbon and/or nitrogen\cite{naguib2011two}. Similar to their bulk counterparts, MXene monolayers (M$_{n+1}$X$_n$) exhibit common properties like metallicity, high selectivity to intercalants, and a high elastic constant (c$_{11}$). However, their metallic nature and lack of bandgap limit their application in semiconducting and optical nanodevices\cite{lee2014tunable}. In the chemical etching processes, various functional groups (F, OH, O, or H) attached to both sides of the MXene surface, resulting in M$_{n+1}$X$_n$T$_x$, where T$_x$ represents the terminating groups. While majority of the functionalized MXenes remain metallic, six members exhibit semiconducting behavior with bandgap values ranging from 0.24 to 1.8 eV:  Ti$_2$CO$_2$, Zr$_2$CO$_2$, Hf$_2$CO$_2$, Sc$_2$CO$_2$, Sc$_2$C{(OH)$_2$}, and Sc$_2$CF$_2$\cite{khazaei2013novel}.

To broaden the use of 2D materials, their properties can be further modified by forming heterostructures. Van der Waals (vdW) heterostructures, which involve vertically stacking two different 2D materials bound by weak vdW interactions, are attracting significant attention. Their distinctive features include an extensive interface area between the constituent materials, atomic-level thinness, and the absence of dangling bonds, highlighting their unique attributes\cite{geim2013van}. The interlayer quantum coupling in 2D heterostructures gives rise to new fascinating phenomena, which will lead to revolutionary electronics and optical applications, such as tunneling and bipolar transistors\cite{roy2015dual,lopez2014light}, as well as flexible optoelectronic devices\cite{huo2014novel}. Semiconducting heterostructures are categorized based on their band alignments into three types: type-I, type-II, and type-III\cite{marschall2014semiconductor}. In type-I heterostructures, the valence band maximum (VBM) and the conduction band minimum (CBM) reside in the same layer. Type-II heterostructures (staggered gap) have the CBM and VBM in different monolayers, whereas type-III heterostructures (broken gap) exhibit no band gap\cite{ozccelik2016band}. Heterostructures offer new ways to combine the good qualities of the materials they are made of\cite{deng2016k, novoselov20162d, ceballos2014ultrafast,arora2021interlayer}. Researchers are particularly interested in semiconducting heterostructures due to their intriguing electronic structures and optical properties, which are promising for applications in photocatalysis, photovoltaics, and optoelectronics. The type-II heterostructures have demonstrated utility in light detection and harvesting applications\cite{kang2013band,wei2014modulating}. Such band alignment substantially enhances electron-hole separation efficiencies while concurrently diminishing charge recombination probabilities, thus highly desirable for photovoltaic and photocatalytic devices\cite{peng2016electronic}. Forming heterostructures enhances mechanical, electronic, optical and thermoelectric properties of MXenes. By combining MXenes with monolayers of other 2D materials, adhering to a lattice parameter mismatch criterion of less than 5$\%$, substantial improvements can be achieved. Notably, the formation of the MXene/TiO$_2$ \cite{xu2021interfacial} interface, characterized by substantial charge transfer and strong adhesion, offers valuable insights for customizing these structures to meet specific needs in electrical energy storage applications. Furthermore, the MXene/Graphene heterostructure enhances electrical conductivity, Li adsorption, and mechanical stiffness\cite{du2018mxene}, while MXene/TMDs heterostructures show promise for gas sensors detecting toxic gases\cite{zhou2023dft}.

Single-layer buckled blue phosphorene (blueP) having a flatter layer than puckered black phosphorene (BlackP), has attracted intense research interest \cite{li2019electronic} due to exceptional qualities, including sizable bandgap and ultra-high mobility. The electronic and magnetic properties of blueP are highly sensitive to foreign adatoms. For instance, a spin-gapless semiconducting state can result from Co and Ag adatoms, while a half-metallic state can be achieved through the adsorption of N and P\cite{ding2015structural}. Band gap engineering of blueP under external strain and electric fields is of great interest for applications in nanoelectronics and rechargeable Li-ion batteries\cite{li2015theoretical}. Due to their similar crystal structure and comparable lattice constant, blueP is an excellent material to practically construct vdW heterostructures with 2D M$_2$CO$_2$ (M= Sc,Zr,Hf). Various  combinations of MXene/blueP heterostructures have been designed theoretically and studied for applications such as water-splitting photocatalysts \cite{Li2020} and as electrode materials for metal ion batteries \cite{Yuan2022}. Enhanced optical absorption has been reported in blueP/Sc$_2$CX$_2$ (X= O, F, OH) vdW heterostructures compared to their parent materials, indicating their potential for optoelectronics and photovoltaic device applications\cite{rehman2019van}.

For the integration of these heterostructures in devices, it is crucial to understand their charge transfer as well as the mechanical stability and tunability of their materials properties under biaxial and uniaxial strain along different direction. Biaxial strain, for instance, applies uniform stress in two dimensions while uniaxial strain, on the other hand, applies stress along a single direction, enabling the manipulation of material anisotropy, bandgap and electronic states, crucial for nanoelectronics and optoelectronic applications. A theoretical study on Zr$_2$CO$_2$/blueP and Hf$_2$CO$_2$/blueP heterostructures have shown type-I to type-II transition under biaxial strain \cite{guo2017strain}. Their findings show that the maximum biaxial strains that can be sustained are 0.16 for blueP, 0.18 for Zr$_2$CO$_2$, 0.19 for Hf$_2$CO$_2$, 0.16 for Zr$_2$CO$_2$/blueP, and 0.17 for Hf$_2$CO$_2$/blueP. Since type-I heterostructures are quite common\cite{ozccelik2016band}, this transition to type-II is significant for the design and application of optoelectronic devices\cite{huang2023dft}.

Many 2D materials show potential applications as thermoelectric (TE) materials offering a promising avenue for converting waste heat into electricity\cite{zhu2021bottom, zhao2021modification}. Thermoelectric generators have become very important devices as we navigate rapid technological advancements and have been facing energy challenges.  With applications spanning decades in aerospace and automotive sectors, particularly in high-temperature environments\cite{jia2021suppressing}, TE generators have garnered substantial interest. Their recent integration into wearable electronics\cite{hong2019wearable} underscores their expanding utility. The application of strain to enhance the power factor and figure of merit (ZT$_e$) of various materials has been well-documented\cite{kumari2023strain, sun2024enhancement}. MXenes are emerging as promising candidates for thermoelectric applications due to their cost-effectiveness, environmental benignity, and potential for high activity\cite{abid2021structural}. BlueP possesses a moderate band gap, lower effective mass, and thereby high carrier mobility, which in turn contributes to an elevated Seebeck coefficient and enhanced electrical conductivity\cite{liu2018first}. Since heterostructure combinations have the potential to enhance charge carrier separation and reduce recombination, it is important to investigate the thermoelectric properties of MXene-blueP heterostructures in detail under strain for their application in TE devices.

 In this investigation, employing first-principles calculations, we predict optimized stacking and geometries of three MXene/blueP heterostructures i.e., Hf$_2$CO$_2$/blueP, Zr$_2$CO$_2$/blueP and Sc$_2$CO$_2$/blueP and present in details about their electronic properties, charge redistribution and stability.  We extend our investigation to the profound influence of strain (biaxial and uniaxial strain) on the intrinsic properties of the M$_2$CO$_2$ (M= Sc,Zr,Hf)/blueP heterostructures. Given that strain has previously demonstrated its capacity to augment the thermoelectric performance of materials, we have conducted calculations on various thermoelectric coefficients, focusing on enhancing the power factor. Our findings will be helpful for researchers and device engineers to apply these materials for applications in various devices.

\section{COMPUTATIONAL DETAILS}
In this work, all the calculations were executed utilizing \textit{ab initio} Density Functional Theory (DFT) in conjunction with all-electron projected augmented wave potentials (PAW)\cite{kresse1999ultrasoft}. The Perdew-Burke-Ernzerhof generalized gradient approximation (PBE-GGA)\cite{perdew1996generalized} was employed for electronic exchange and correlation, as incorporated within the Vienna \textit{Ab initio} Simulation Package (VASP)\cite{kresse1996efficient} with periodic boundary conditions. The optB88-vdW functional\cite{klimevs2011van} was integrated to account for vdW interactions. A non-local correlation energy was incorporated in addition to the semilocal exchange-correlation energy. It was necessary to make adjustments to get accurate contributions from the atom's adjacent spheres, given that vdW density functionals generally yield densities that are less spherical compared to standard GGA functionals. The plane wave basis set was converged using an energy cut-off of 500 eV. A vacuum of 24 ${\mathrm{\mathring{A}}}$
was applied to minimize the interaction of the heterostructure slab and its periodic images along the z direction. The
Brillouin zone was sampled with a 11 $\times$ 11 $\times$ 1 k mesh within the
Monkhorst-Pack scheme. The convergence criteria of $10^{-4}$ eV for the total energy and 0.01 eV/$\si{\angstrom}$ for force were used for the structural optimization.            
The formation energy ($E_f$) of the interface between M$_2$CO$_2$ and blueP was calculated as follows: 
\begin{equation}
    E_{f} = (E_{Total}-(E_{MXene} +E_{blueP}) )
\end{equation}
where, $E_{Total}$, $E_{MXene}$ and $E_{blueP}$ represent the energies of the heterostructure, the MXene layer, and blueP layer, respectively. 

The charge density difference ($\Delta\rho$) was calculated by, 
\begin{equation}
    \Delta\rho = \Delta\rho_{MXene/blueP}-\Delta\rho_{MXene}-\Delta\rho_{blueP}
\end{equation}
where, $\Delta\rho_{MXene/blueP}$, $\Delta\rho_{MXene}$ and $\Delta\rho_{blueP}$ represent the total charge density of the MXene/blueP heterostructures, pristine MXene, and blueP monolayers, respectively.

Phonon spectra calculations were carried out using the Phonopy\cite{togo2015first} code interfaced with VASP, utilizing harmonic interatomic force constants obtained through density functional perturbation theory (DFPT)\cite{PhysRevLett.58.1861,PhysRev.139.A796}. Supercells of 4$\times$4$\times$1 were employed for these calculations. Thermal stability calculations were conducted at 300 K utilizing $\textit{Ab Initio}$ Molecular Dynamics (AIMD) simulations\cite{car1985unified} with the VASP code. Canonical ensemble was employed for molecular dynamics (MD) simulations to achieve atomic structure thermal equilibration. The Noose-Hoover thermostat\cite{evans1985nose}, as implemented in the VASP package, was utilized within this ensemble including a time step of 1 fs for 3000 steps. Electronic band structures and density of states (DOS) were plotted using the open-source Python package sumo\cite{ganose2018sumo}. Visualization of atomic structures and charge density was plotted using the software VESTA\cite{momma2011vesta}.

The thermoelectric properties for all the heterostructures under strain were computed utilizing semi-classical Boltzmann transport theory within the updated Boltztrap2 code\cite{BoltzTraP2}, leveraging the electronic structure derived from first-principles calculations. Given that the TE coefficients are significantly dependent on carrier concentration, a rigid band approximation is employed to simulate doping effects. This method presumes that variations in temperature or doping levels do not alter the band structure. We selected a temperature of 900K to focus on high-temperature applications of the material under investigation. A narrower range of charge carrier concentrations was opted for due to the ease of achieving such minimal doping levels in experiments. We have employed the linearized version of the Boltzmann Transport Equation (BTE) under the Constant Scattering Time Approximation (CSTA), where the transport distribution function is given by;
\begin{equation}
    \sigma(\varepsilon, T)=\int \sum_b \mathbf{v}_{b, \mathbf{k}} \otimes \mathbf{v}_{b, \mathbf{k}} \tau_{b, \mathbf{k}} \delta\left(\varepsilon-\varepsilon_{b, \mathbf{k}}\right) \frac{\mathrm{d} \mathbf{k}}{8 \pi^3}
\end{equation}
where, $\varepsilon$ is the energy of charge carriers, \textbf{k} is the wave vector, \textbf{v} is the group velocity of the charge carriers, \textit{b} is the index related to energy bands and $\tau$ is the relaxation time of charge carriers. This is employed to calculate generalized transport coefficient of order $\alpha$, as a function of chemical potential $\mu$ and temperature T as;
\begin{equation}
    \mathcal{L}^{(\alpha)}(\mu ; T)=q^2 \int \sigma(\varepsilon, T)(\varepsilon-\mu)^\alpha\left(-\frac{\partial f^{(0)}(\varepsilon ; \mu, T)}{\partial \varepsilon}\right) \mathrm{d} \varepsilon
\end{equation}
where q is the electronic charge, and $f^{(0)}$ is the Fermi-Dirac distribution function. This further gives the charge and heat currents as;
\begin{equation}
    \begin{aligned}
& j_e=\mathcal{L}^{(0)} \mathbf{E}+\frac{\mathcal{L}^{(1)}}{q T}(-\nabla T) \\
& j_Q=\frac{\mathcal{L}^{(1)}}{q} \mathbf{E}+\frac{\mathcal{L}^{(2)}}{q^2 T}(-\nabla T) 
\end{aligned}
\end{equation}
By considering the two experimental conditions of zero temperature gradient and zero electric current, we can determine the electrical conductivity, Seebeck coefficient, and the electronic contribution to the thermal conductivity as;
\begin{equation}
    \begin{aligned}
\sigma & =\mathcal{L}^{(0)} \\
S & =\frac{1}{q T} \frac{\mathcal{L}^{(1)}}{\mathcal{L}^{(0)}} \\
\kappa_e & =\frac{1}{q^2 T}\left[\frac{\left(\mathcal{L}^{(1)}\right)^2}{\mathcal{L}^{(0)}}-\mathcal{L}^{(2)}\right]. 
    \end{aligned}
\end{equation}

\section{RESULTS AND DISCUSSION}

\subsection{Structural and electronic properties of the monolayers}

\begin{figure}[h!]
	\centering
	\includegraphics[width=\columnwidth]{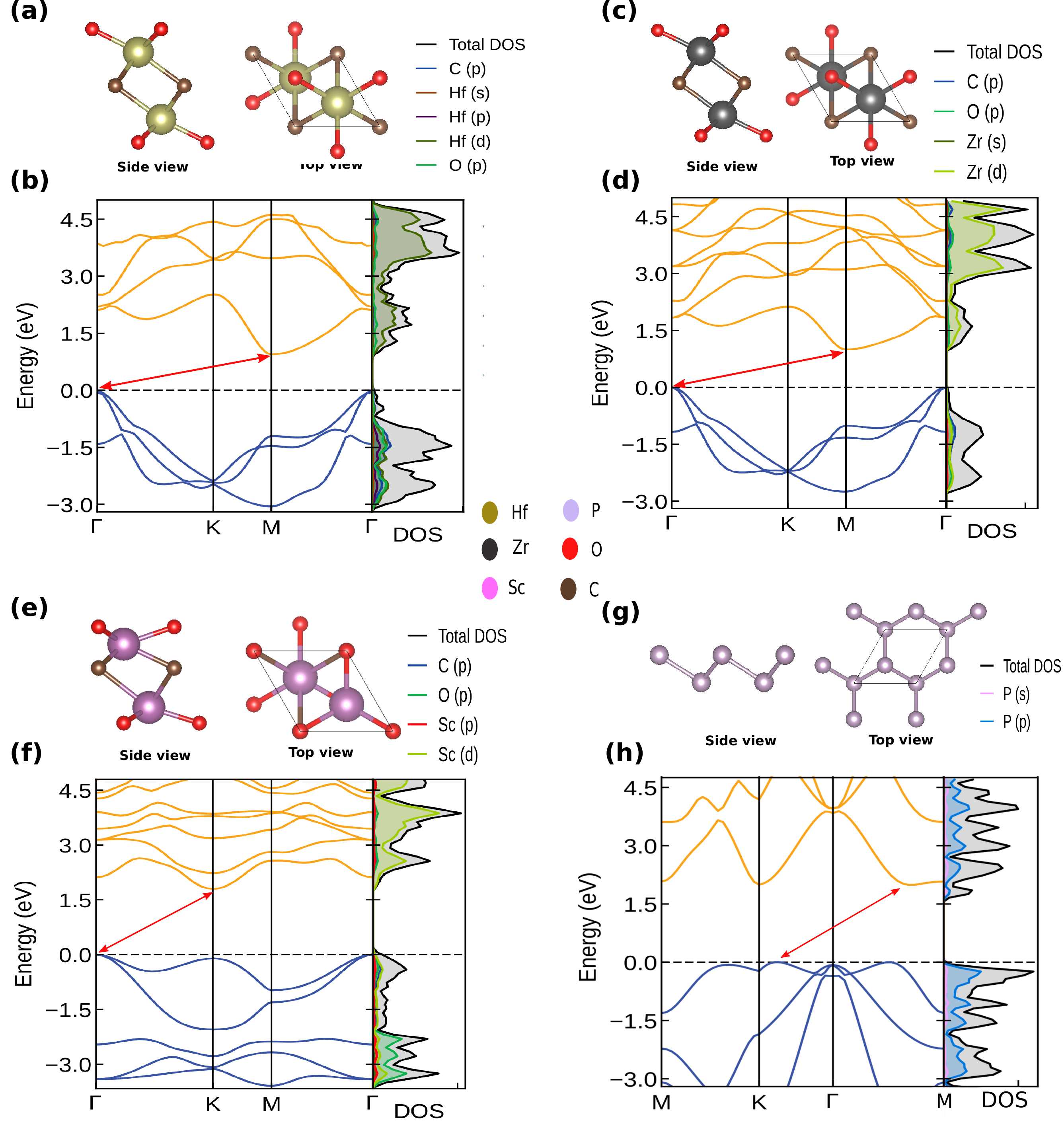}
	\caption{Crystal structure of MXenes M$_2$CO$_2$, (a) Hf$_2$CO$_2$, (c) Zr$_2$CO$_2$, (e) Sc$_2$CO$_2$ and (g) blueP and Band structure of MXenes M$_2$CO$_2$\textcolor{red}{,} (b) Hf$_2$CO$_2$, (d) Zr$_2$CO$_2$, (f) Sc$_2$CO$_2$ and (h) blueP, respectively. The position of the VBM and CBM is illustrated by a double-sided red arrow }
	\label{fig1}
\end{figure}

We first optimized the lattice structures of the constituent monolayers of the heterostructures and then calculated their electronic properties. The most stable optimized structures, band structures, and DOS of M$_2$CO$_2$ (M = Sc, Hf, and Zr) and blueP monolayers are shown in Figure  \ref{fig1}. In Zr$_2$CO$_2$ and Hf$_2$CO$_2$, the O atoms in the upper (bottom) atomic layer of the MXene structure are positioned directly above (below) the transition metal atoms \cite{gandi2016thermoelectric}. Due to this structural and compositional similarity, Zr$_2$CO$_2$ and Hf$_2$CO$_2$ have many similar properties. On the other hand, Sc$_2$CO$_2$ has different sites for the O atoms on the two sides: one functional O layer is positioned at the hollow site of three neighboring C atoms, and the other O atom points to a C atom (Figure \ref{fig1} (e)), making its configuration asymmetric along the thickness direction. The top view of the blueP structure is similar to the honeycomb lattice of graphene and has two covalently bonded P atoms alternatingly being displaced out of the 2D plane with an equilibrium distance of 2.26 ${\mathrm{\mathring{A}}}$ between them\cite{li2019electronic}. 
The optimized lattice parameters (L$_0$), band gaps (E$_g$) and work functions ($\Phi$) of the constituent materials of the heterostructure calculated using GGA-PBE functional are given in Table \ref{table1}, while the bond lengths are presented in Table S1 of the Supporting Information.

\begin{table}[h]
	\begin{tabular}{|l|c|l|c|}
	\hline
\textbf{Monolayer} & \textbf{L$_0$ (\si{\angstrom})} &  \textbf{E$_g$ (eV)}  & \textbf{$\Phi$ (eV)}\\ \hline
\hspace{3mm} Hf$_2$CO$_2$  & 3.26 &  \hspace{4mm}0.94 & 5.29 \\ \hline
\hspace{3mm} Zr$_2$CO$_2$& 3.31 &  \hspace{4mm}0.99 & 5.18\\ \hline
\hspace{3mm} Sc$_2$CO$_2$& 3.45 & \hspace{4mm}1.74  & 5.35\\ \hline
\hspace{3mm} blueP & 3.29 & \hspace{4mm}1.99 & 6.09\\ \hline
\end{tabular}
	\caption{Lattice parameters (L$_0$), band gaps (E$_g$) and work functions ($\Phi$) of the constituent materials of the heterostructure calculated using GGA-PBE functional}
	\label{table1}
\end{table}

The respective band structures and DOS of these monolayers are shown in Figure \ref{fig1} (b,d,f,h). Zr$_2$CO$_2$ and Hf$_2$CO$_2$ exhibit similar band dispersion, with indirect band gaps of 0.99 eV and 0.94 eV, respectively. For both materials, the VBM is at the $\Gamma$ point, and the CBM is at the M point of the Brillouin zone. The conduction band edge is formed by M-d and O-p states, while the valence band edge is composed of C-p, O-p, and M-d states. Sc$_2$CO$_2$ is also an indirect band gap semiconductor, with a band gap of 1.74 eV. However, its band dispersion differs, with the VBM (CBM) is located at $\Gamma$ (K) point. The DOS plot reveals that C-p, O-p, and Sc-d states dominate the valence band, while the Sc-d state is dominant in the conduction band. The C-p and Sc-d states are strongly hybridized, which is consistent with the fact that MXenes are mainly assembled through the bond between C and the transition metal atom. The band structure of blueP shows the location of VBM  in between K and $\Gamma$ point and CBM  in between $\Gamma$ and  M point giving an indirect band gap of 1.99 eV.  Both the valence and conduction bands are predominantly derived from P-p states.

\subsection{MXenes (M$_2$CO$_2$) and blueP heterostructures}

 The lattice mismatches between M$_2$CO$_2$ (M= Hf, Zr, Sc) and blueP are 0.91$\%$, 0.60$\%$ and 4.6$\%$\textcolor{red}{,} respectively, which all are within the acceptable range of 5\%  and their interface is flexible enough to accommodate this lattice mismatch, indicating that these heterostructures can be synthesized experimentally\cite{li2020blue}. We have considered 6 different stackings of MXene/blueP heterostructures to determine the most stable configuration. These configurations are all properly optimized. Figure S1 and Figure S2 in the Supporting Information illustrate the stacking configurations considered for the (M = Hf, Zr)$_2$CO$_2$/blueP and Sc$_2$CO$_2$/blueP vdW heterostructures, respectively. To verify the thermodynamic stability of the most favorable stacking within this set of configurations, we have computed the formation energy. The calculated formation energies for different stackings are listed in Table S2 in the Supporting Information. It is noteworthy that all configurations exhibit negative formation energies, signifying the energetically viable nature of M$_2$CO$_2$/blueP heterostructure fabrication. In accordance with the fundamental principle that a more negative formation energy indicates greater stability,  we have considered the most stable structure for our further study, as indicated in bold in Table S2 in the Supporting Information. The optimized interlayer distance for these most stable stackings are found to be 2.94${\mathrm{\mathring{A}}}$, 2.89${\mathrm{\mathring{A}}}$ and 2.50${\mathrm{\mathring{A}}}$ for Hf$_2$CO$_2$/blueP, Zr$_2$CO$_2$/blueP and Sc$_2$CO$_2$/blueP, respectively. These values match well with the previous reports for these heterostructures \cite{rehman2019van,guo2017strain}.
 
\begin{figure}[h!]
	\centering
	\includegraphics[width=\columnwidth]{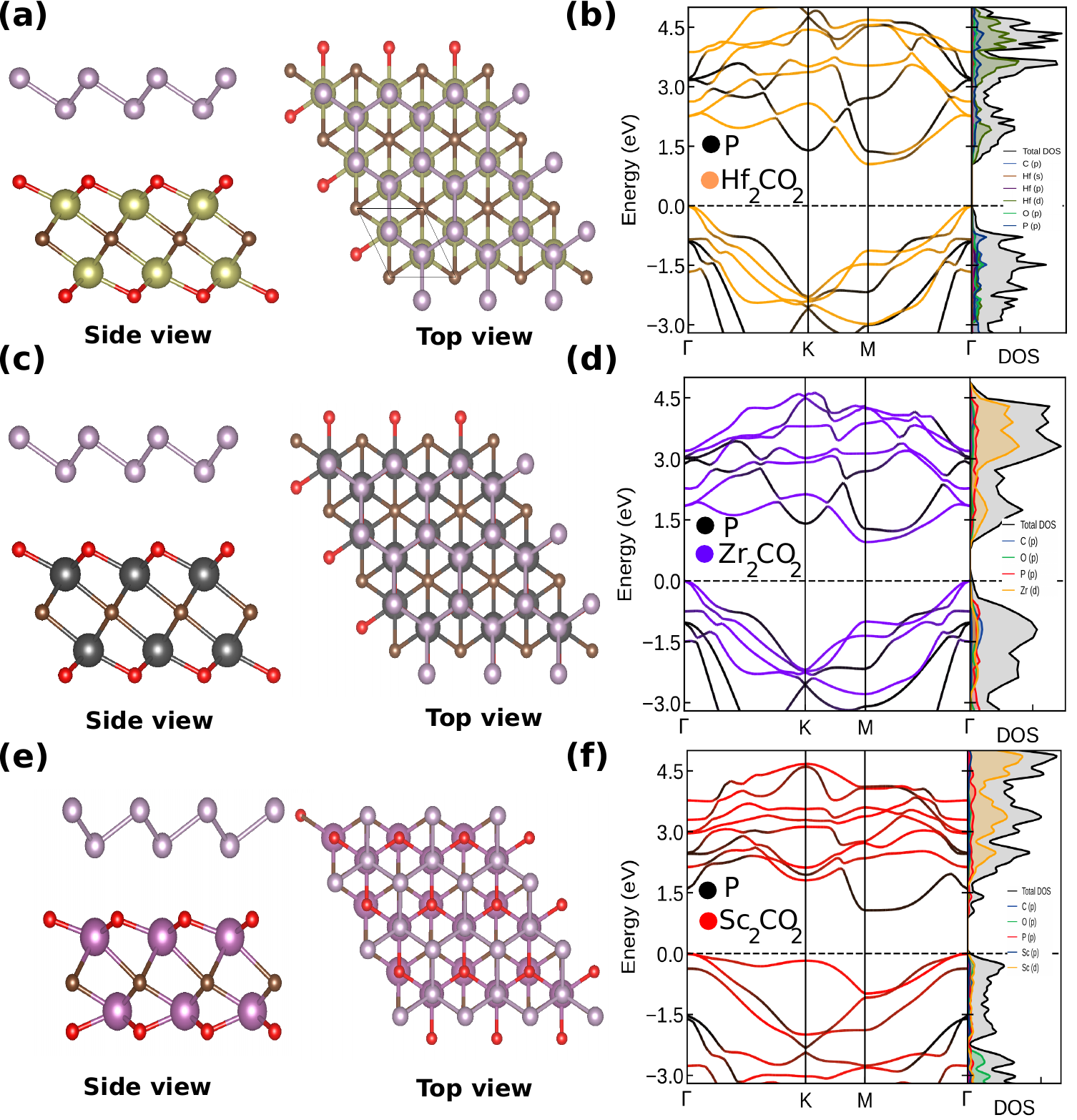}
	\caption{Crystal structure of MXene/blueP heterostructures (a) Hf$_2$CO$_2$/blueP, (c) Zr$_2$CO$_2$/blueP, and (e) Sc$_2$CO$_2$/blueP and Band structure of (b) Hf$_2$CO$_2$/blueP\textcolor{red}{,} (d) Zr$_2$CO$_2$/blueP, and (f) Sc$_2$CO$_2$/blueP\textcolor{red}{,} respectively}
	\label{fig2}
\end{figure}

The atomic structure, band structure, and DOS of M$_2$CO$_2$ (M=Hf, Zr, Sc)/blueP heterostructures are shown in Figure \ref{fig2}. The results reveal that Hf$_2$CO$_2$/blueP, Zr$_2$CO$_2$/blueP, and Sc$_2$CO$_2$/blueP all exhibit indirect band gaps, with values of 1.04 eV, 0.95 eV, and 1.06 eV, respectively. In each case, the VBM is located at the $\Gamma$ point, while the CBM is at the M point. The projected band structures indicate that in both Hf$_2$CO$_2$/blueP and Zr$_2$CO$_2$/blueP, the CBM and VBM are predominantly influenced by the MXene layer. Specifically, the VBM is primarily determined by C-p and O-p states, while the CBM is influenced by the d states of Hf (or Zr) atoms, classifying these as type-I heterostructures. There is a small reduction of band gap in Sc$_2$CO$_2$/blueP as compared to pristine layers suggesting significant inter-layer interactions and charge redistribution in this heterostructure. BlueP has more contribution in the CBM (P-p states),  while the VBM (Sc-d and O-p states) is mainly contributed from Sc$_2$CO$_2$ monolayer, indicating a type-II band alignment.

\begin{figure}[h!]
	\centering
	\includegraphics[width=\columnwidth]{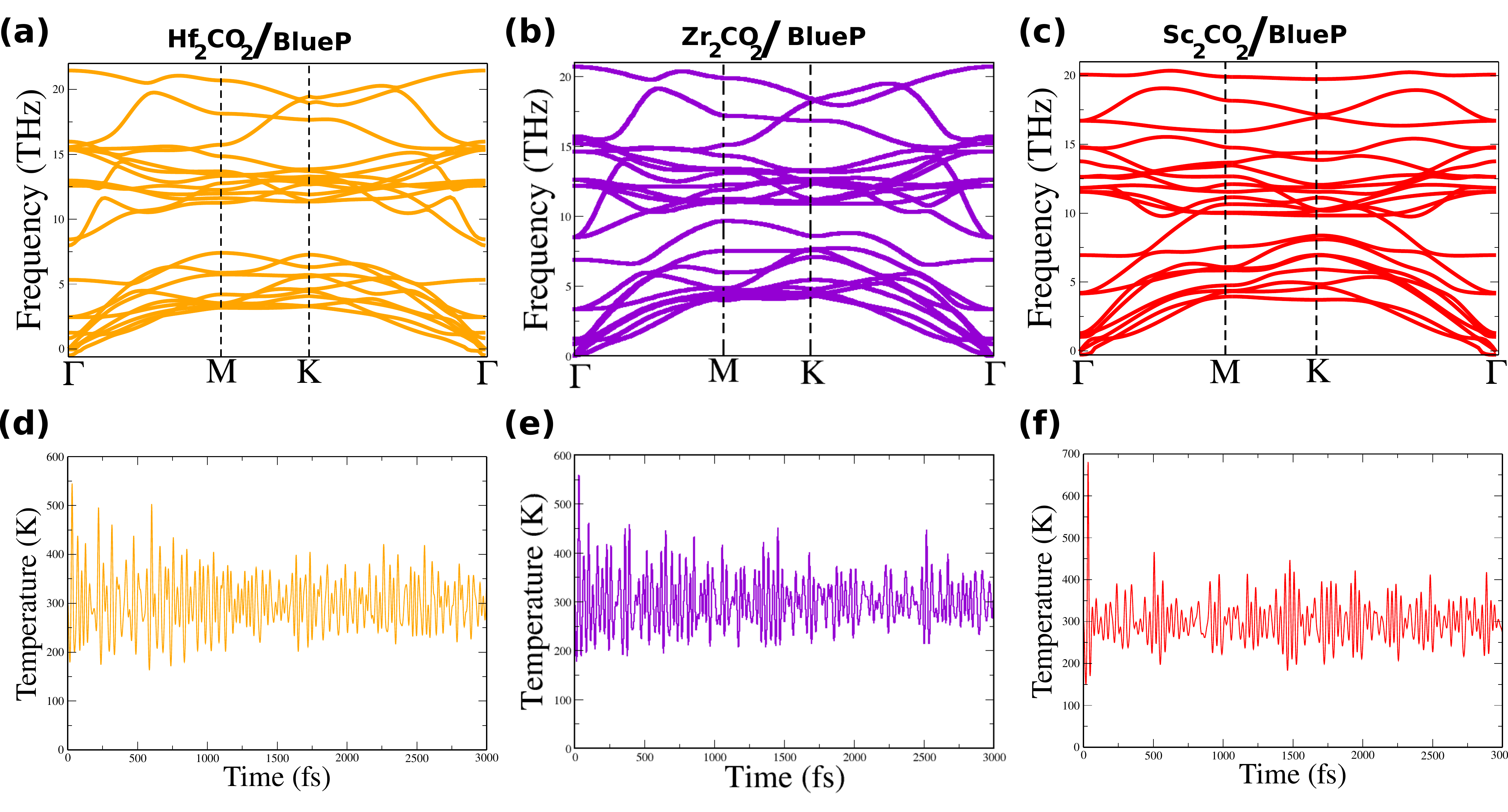}
	\caption{Phonon band structure of heterostructures, (a) Hf$_2$CO$_2$/blueP, (b) Zr$_2$CO$_2$/blueP, and (c) Sc$_2$CO$_2$/blueP and Thermal stability at 300 K of heterostructures (d) Hf$_2$CO$_2$/blueP, (e) Zr$_2$CO$_2$/blueP, and (f) Sc$_2$CO$_2$/blueP, respectively }
	\label{fig3}
\end{figure}

To verify the dynamic stability of the heterostructures, we performed phonon spectrum calculations, as depicted in Figure \ref{fig3} (a-c). The absence of imaginary frequencies in the phonon spectra confirms the dynamical stability of the MXene/blueP heterostructures. Although there is a small U-shaped feature near the $\Gamma$ point in the phonon spectra, it does not indicate lattice instability. Instead, this feature represents the flexural acoustic mode characteristic of 2D systems \cite{singh2017giant}. To further assess the thermal stability of the M$_2$CO$_2$/blueP heterostructures, we conducted \textit{ab initio} molecular dynamics (AIMD) simulations using the Nose-Hoover heat bath scheme \cite{nose1984unified}. The simulations were performed on a 3$\times$3$\times$1 supercell over a period of 3000 fs at 300 K. The temperature evolution throughout the simulation is shown in Figure \ref{fig3} (d-f). The observed convergence of temperature at the end of the simulation steps, affirms the robustness and accuracy of the obtained results. These results affirm the thermal stability of the considered heterostructures at room temperature (300 K), rendering them viable candidates for future nanoelectronic applications.

In addition, the mechanical stability of all considered monolayers and heterostructures has been verified. The elastic constants were calculated to evaluate the mechanical stability of blueP, Hf$_2$CO$_2$, Zr$_2$CO$_2$, Sc$_2$CO$_2$, and their corresponding heterostructures (Hf$_2$CO$_2$/blueP, Zr$_2$CO$_2$/blueP, and Sc$_2$CO$_2$/blueP). Since these materials exhibit an isotropic hexagonal lattice, the elastic tensor components are reduced to three key constants: C$_{11}$, C$_{12}$, and C$_{66}$ (with C$_{66}$ = (C$_{11}$ - C$_{12}$)/2), where C$_{ij}$ represents the second-order elastic constant, in unit of N/m, for two-dimensional materials\cite{gao2023mechanical}. Only two independent elastic constants, C$_{11}$ and C$_{12}$, are required to describe the system. These constants were obtained through the stress-strain relationship by applying a minor strain to the equilibrium lattice configuration. The values of  C$_{11}$, C$_{12}$, C$_{66}$  meet the mechanical stability criteria (C$_{11} > 0$ and $C_{11} > |C_{12}|$) , as shown in supplementary Figure S3, confirming the mechanical stability of both the monolayers and heterostructures.

The work function is a key parameter that describes a material's ability to release electrons from its surface. It is defined as the energy required for an electron to transition from the Fermi level to the vacuum level. The work function $(\Phi)$ can be calculated as the difference between the vacuum ($E_{vac} $) and the Fermi level (E$_{F}$). Mathematically, it is expressed as;
\begin{equation}
    \Phi =E_{{vac} } - E_{F}
\end{equation}
\begin{figure}[h!]
	\centering
	\includegraphics[width=\columnwidth]{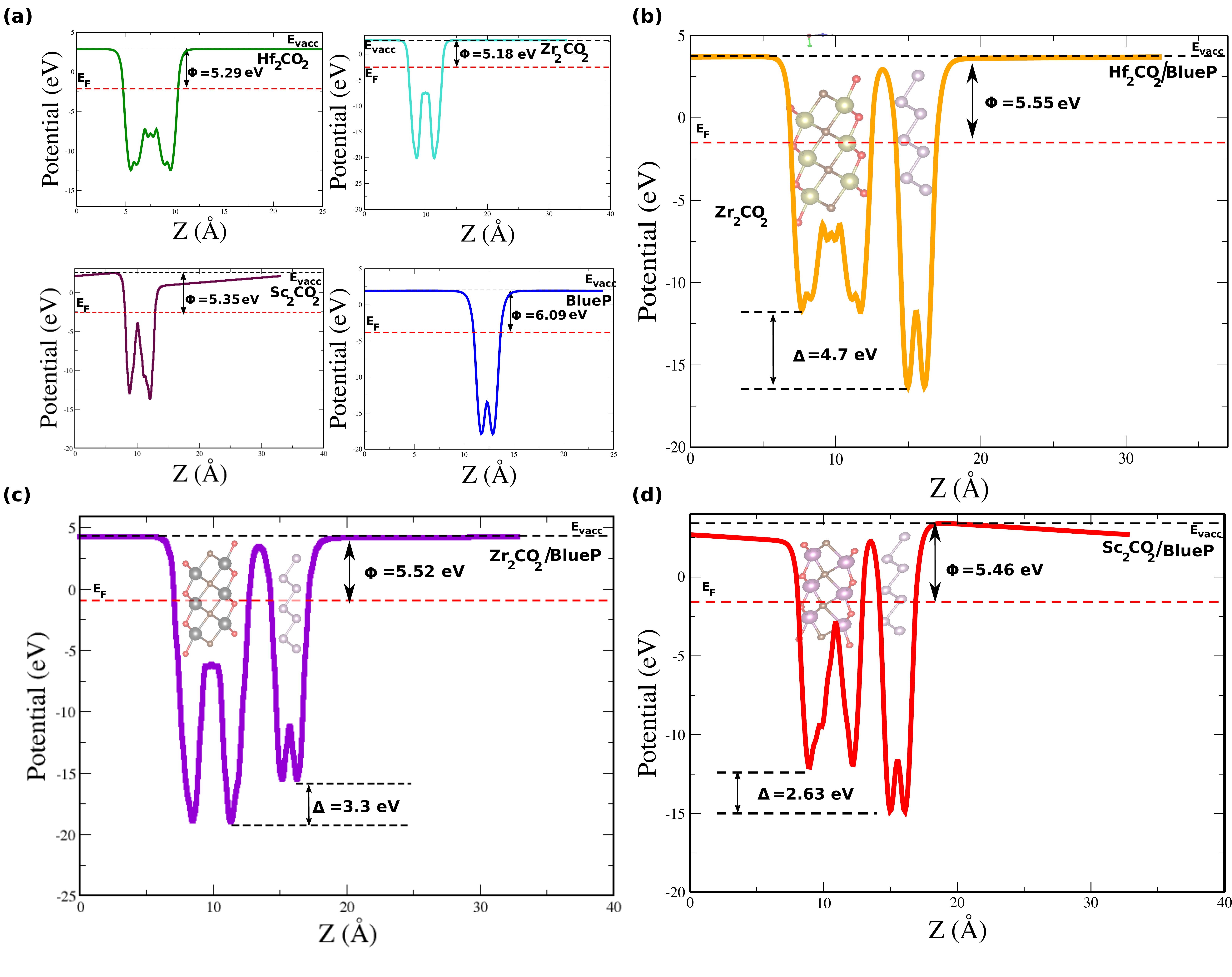}
	\caption{(a) Electrostatic Potential of monolayers and Electrostatic Potential of heterostructures (b) Hf$_2$CO$_2$/blueP, (c) Zr$_2$CO$_2$/blueP, and (d) Sc$_2$CO$_2$/blueP, respectively. The black dotted line represents  E$_{vacc}$, while the red dotted line indicates E$_{Fermi}$ }
	\label{fig4}
\end{figure}
  The energy level $E_{vac}$ is identified in the electrostatic potentials along the z-direction, as shown in Figure \ref{fig4} by the black dotted line. The  $E_F$ level is obtained from first-principles calculations, indicated by the red dotted line. For Sc$_2$CO$_2$,  its asymmetric atomic structure\cite{khazaei2013novel} induces a spontaneous polarization that results in an internal electric field between the two sides of the monolayer \cite{Lee2014pccp}. As a result, the electrostatic potential shows an asymmetry in the vacuum levels for two surfaces of the Sc$_2$CO$_2$ monolayer. The work function values of the monolayers are tabulated in Table \ref{table1}. Electrons preferentially move from lower to higher work function materials, seeking a lower energy state 
while forming heterostructures. The result indicated that when the monolayer MXene interacted with the monolayer blueP to form the MXene/blueP heterostructure the electrons would flow from the MXene to blueP, regardless of the band alignment and CBM position, until they reached the same Fermi levels. This transfer results in Fermi level adjustments with respect to the vacuum level: an elevation in blueP and a corresponding reduction in MXene, establishing work function values of 5.55, 5.52, and 5.46 for Hf$_2$CO$_2$/blueP, Zr$_2$CO$_2$/blueP, and Sc$_2$CO$_2$/blueP, respectively.

The electrostatic potential energy graph of the heterostructures is shown in Figure \ref{fig4} (b-d). The potential drops across the interfaces are 4.7, 3.3, and 2.63 eV for the Hf$_2$CO$_2$/blueP, Zr$_2$CO$_2$/blueP, and Sc$_2$CO$_2$/blueP heterostructures, respectively. These values indicate the presence of a strong electrostatic field perpendicular to the interface. In type-II heterostructures, this built-in electric field is crucial for controlling the recombination of electron-hole pairs and significantly enhancing the migration of photogenerated charges. Effective separation of these charge carriers is essential for maximizing efficiency in photovoltaic applications, as electron-hole pairs generated by incident photons often recombine within the semiconductor or at the surface before contributing to photocatalytic effects. Type-II band alignment is particularly advantageous, as it facilitates the separation of charge carriers by allowing electrons to migrate to the CBM of one layer while holes remain confined to the VBM of the other layer\cite{huang2023dft}. The built-in electric field in these heterostructures effectively separete the electron-hole pair and reduces charge recombination, significantly boosting the photocatalytic performance of the materials. This improvement in charge separation and transport markedly boosts the photocatalytic performance of these materials. Additionally, the enhanced sensitivity  provided by the built-in electric field is effective in detecting minor changes in environmental conditions, making these heterostructures valuable for sensor applications.

\begin{figure}[h!]
	\centering
	\includegraphics[width=\columnwidth]{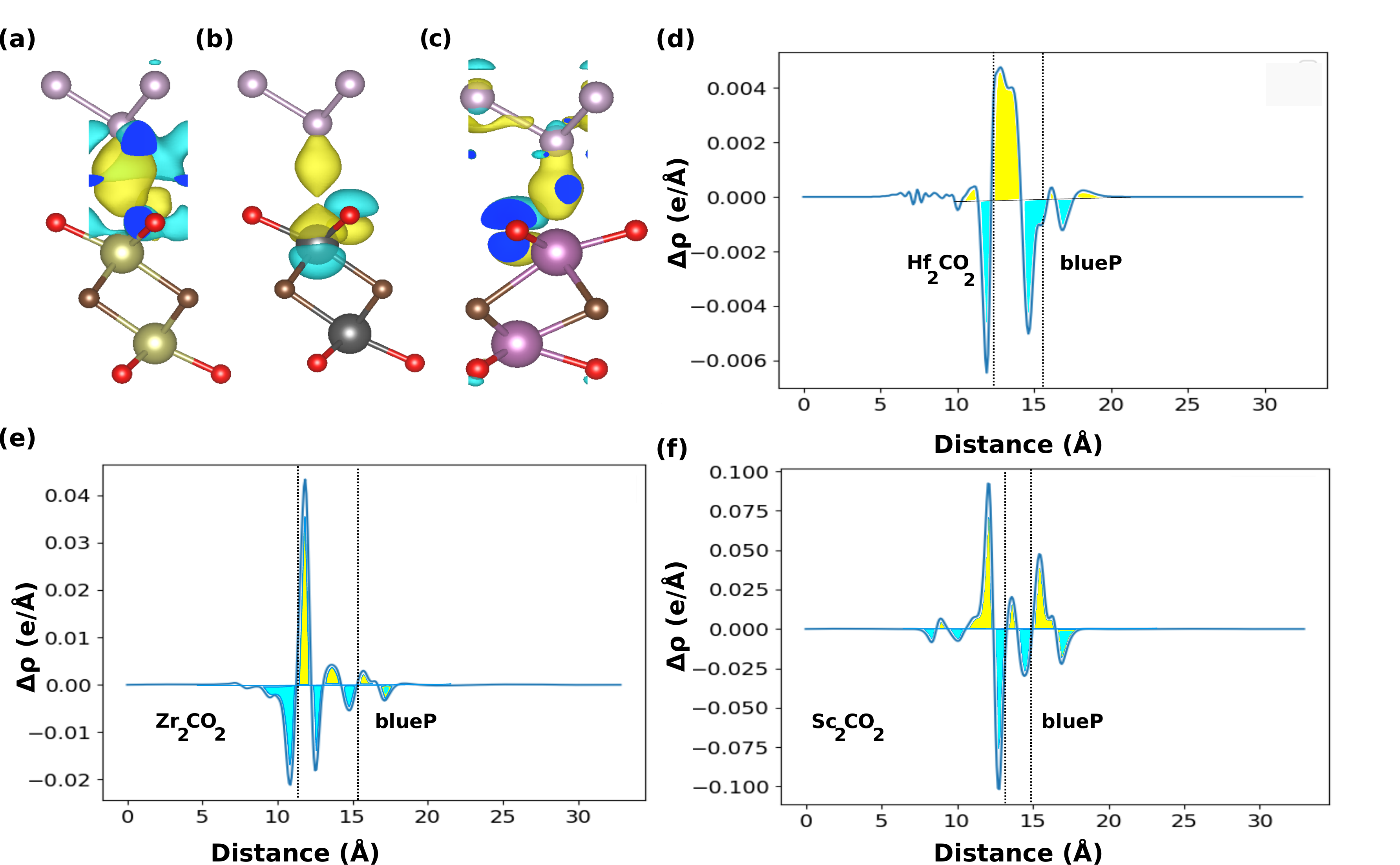}
	\caption{Charge density difference plot of heterostructures, (a) Hf$_2$CO$_2$/blueP, (b) Zr$_2$CO$_2$/blueP, and (c) Sc$_2$CO$_2$/blueP  and Planar charge density of heterostructures (d) Hf$_2$CO$_2$/blueP, (e) Zr$_2$CO$_2$/blueP, and (f) Sc$_2$CO$_2$/blueP respectively. Yellow and cyan
represent charge accumulation and charge depletion, respectively and the area between the dashed lines represents the interface of MXene/blueP heterostructure}
	\label{fig6}
\end{figure}

The formation of a heterostructure induces interactions between the layers, resulting in charge transfer and redistribution at the interface. This phenomenon is examined by calculating the charge density difference of the heterostructure, as shown in Figure \ref{fig6} (a-c). The charge density difference plots are derived by subtracting the charge densities of the individual layers in the unit cell from that of the heterostructure. For a more detailed analysis of the charge transfer processes, the planar charge density difference along the z-direction was also calculated, as depicted in Figure \ref{fig6} (d-f). In the Hf$_2$CO$_2$/blueP and Zr$_2$CO$_2$/blueP heterostructures, charge redistribution is primarily localized around the interface. Since there is no complete charge carrier separation, these two combinations would not be beneficial for photocatalytic applications. In contrast, the Sc$_2$CO$_2$/blueP heterostructure exhibits charge accumulation in the blueP layer, indicating effective charge carrier separation and enhancing its photocatalytic efficiency.
Further charge analysis using the Bader method provides quantitative results regarding charge transfer in the MXene/blueP heterostructures, as summarized in Table S3 in the Supporting Information. From the computed Bader charges, we found that the amount of charge transfer correlates with the interlayer interaction across the interface. Our results show that the charge transfer per unit cell is about 0.0144 $\lvert e \rvert$ in Hf$_2$CO$_2$/blueP, 0.0149 $\lvert e \rvert$ in Zr$_2$CO$_2$/blueP, and 0.0135  $\lvert e \rvert$ in Sc$_2$CO$_2$/blueP heterostructures.

\subsection{Application of external strain on M$_2$(M=Hf, Zr, Sc)CO$_2$/blueP heterostructure}

Nanoelectronics and optoelectronics greatly benefit from materials whose electronic properties, particularly the band gap, can be adjusted by external factors such as strain, electric field, and interlayer distance \cite{li2014elastic}. The physical properties of 2D layered materials are known to be highly sensitive to external influences. In experimental fabrication processes involving 2D materials on a substrate, mismatches between the 2D material and the substrate can lead to crystal distortion, inducing strain or stress effects. Consequently, a comprehensive understanding of the electronic properties of these heterostructures is crucial for optimizing their performance in future applications.\\

\begin{figure}[h!]
	\centering
	\includegraphics[width=\columnwidth]{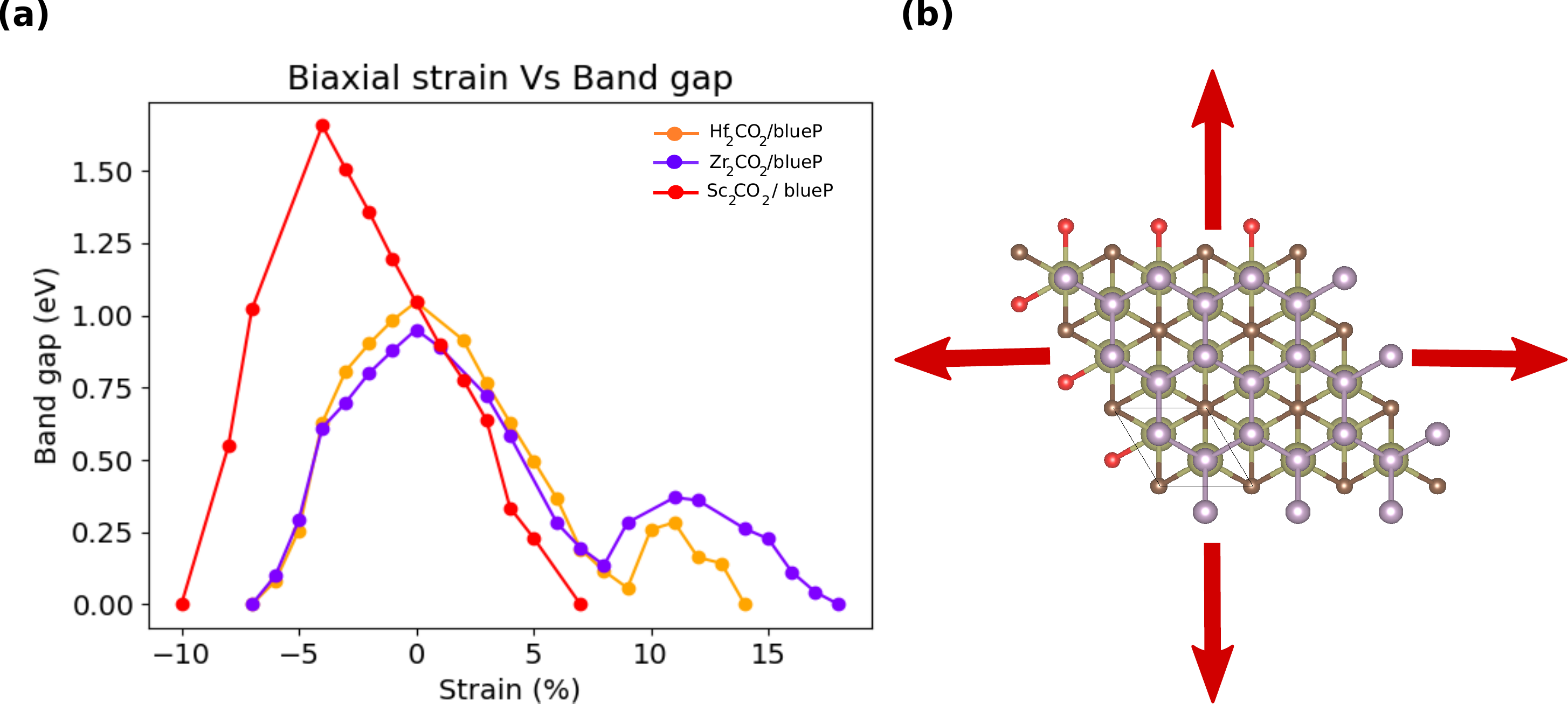}
	\caption{(a) The change in bandgaps of MXene/blueP heterostructures under various percentages of applied biaxial strain and (b) Schematic of biaxial strain}
	\label{fig7}
\end{figure}

{\underline{\textbf{Biaxial strain}}} \textbf{:} First, we applied biaxial strain on the three heterostructures by changing its in-plane lattice parameters along the a and b axis equally. The applied strain was calculated as, $\epsilon = (L-L_0)/L_0$, where L and $L_0$ represent the lattice parameter of the strained and the unstrained structure, respectively. The band gap of the three MXene/blueP heterostructures changes as a function of the biaxial strain, shown in Figure  \ref{fig7}(a). The response of Zr$_2$CO$_2$/blueP and Hf$_2$CO$_2$/blueP to varying strain percentages exhibit remarkable similarities: a discernible reduction in the band gap for both compressive and tensile strain. Both the heterostructures undergo a transition to a metallic state at a compressive strain of 7$\%$.  An increase and decrease in band gap were observed after a tensile strain value of 8$\%$ (9$\%$) in Zr$_2$CO$_2$/blueP (Hf$_2$CO$_2$/blueP), where the band gap attains the first minima. Zr$_2$CO$_2$/BlueP and Hf$_2$CO$_2$/blueP undergo metallic transition at tensile strains of 18$\%$ and 14$\%$, respectively. In contrast, the Sc$_2$CO$_2$/blueP heterostructure demonstrates a distinctive behavior in its band gap response. Initially, the band gap demonstrates an upward trend under compressive strain, then starts decreasing, and ultimately, it transforms into a metallic state at a compressive strain of 10\%. Under tensile strain, the band gap exhibits a sharp decrease, and the material transitions to a metallic state at a strain of 7$\%$.

\begin{figure}[h!]
	\centering
	\includegraphics[width=\columnwidth]{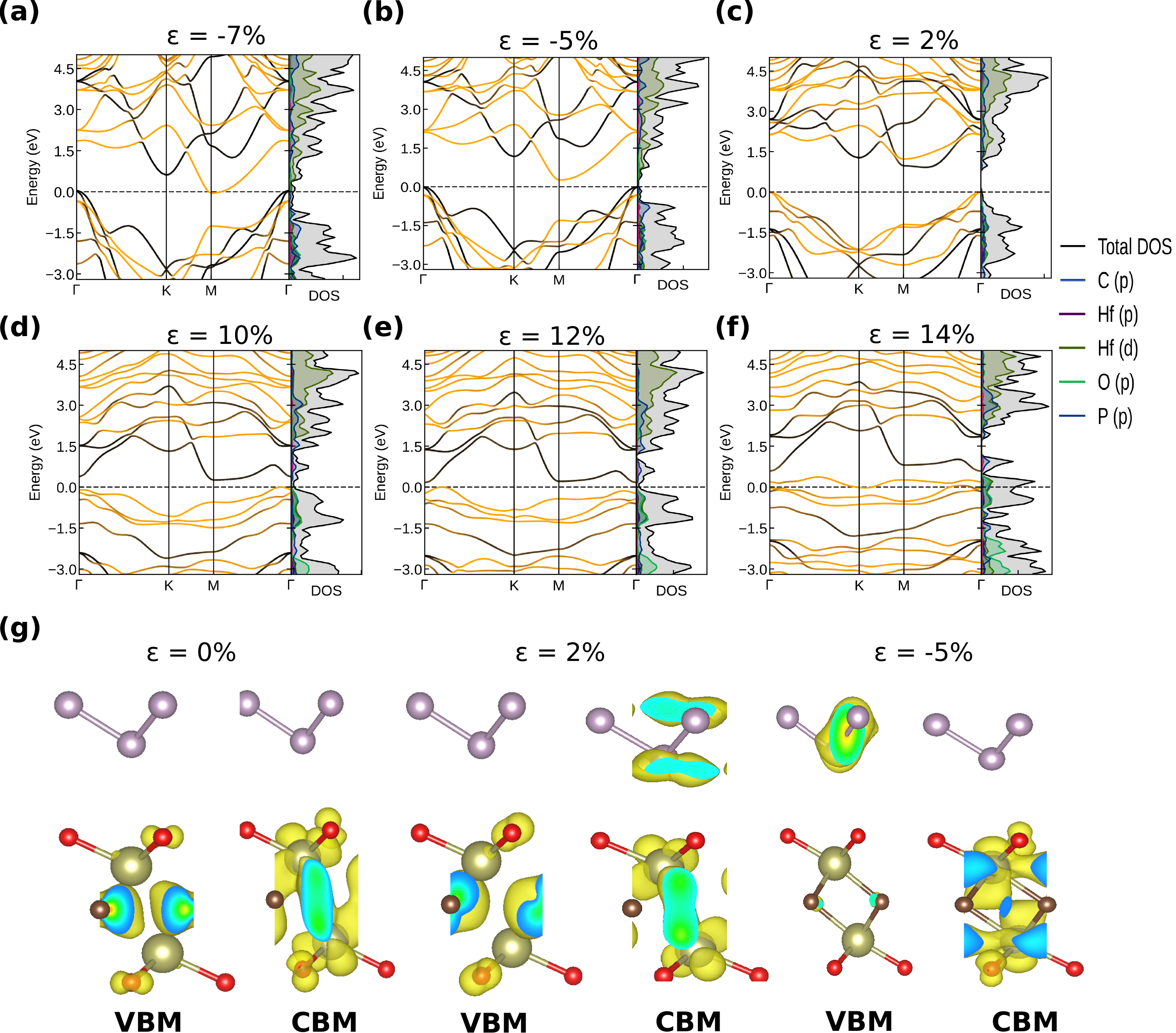}
	\caption{Band structures, PDOS  [(a)-(f)] and (g) Band decomposed charge density of Hf$_2$CO$_2$/blueP heterostructures under different percentages of biaxial strain}
	\label{fig8}
\end{figure}

By examining the band structure and partial density of states (PDOS) of the M$_2$CO$_2$/blueP heterostructure under both compressive and tensile strain, we gained deeper insights into how biaxial strain affects its electronic structure. Figure  \ref{fig8}
shows the bandstructure and PDOS plot of Hf$_2$CO$_2$/blueP under different percentages of biaxial strain. For an unstrained heterostructure of Hf$_2$CO$_2$/blueP, the conduction band is predominantly influenced by the Hf-d state, and the valence band is dominated by C-p and O-p states. The VBM is located at the $\Gamma$ point and the CBM at the M point. As we increase the compressive strain, the CBM at the M point decreases and the VBM at the $\Gamma$ point increases, crossing the fermi level at -7$\%$, resulting in a metallic transition. In the case of tensile strain, the band gap is reduced till 9$\%$, where CBM  decreases at M point with strain. At 10$\%$, the CBM shifts slightly upward, causing the band gap to increase slightly. With further increase in strain, the CBM shifts from the M point to the $\Gamma$ point at 12$\%$ strain, and the VBM also shifts away from the $\Gamma$ point. A semiconductor-to-metal transition (S-M transition) occurs when the CBM crosses the Fermi level at 14$\%$ strain. Under increasing compressive strain, the contribution of the P-p orbitals to the valence band increases, while the conduction band remains dominated by the C-p and O-p orbitals (see Figure  \ref{fig8} (b)), leading to a transition from type-I to type-II. To gain visual insight, we performed band-decomposed charge density calculations as depicted in Figure  \ref{fig8} (g). Our analysis revealed that CBM and VBM contributions of Hf$_2$CO$_2$/blueP originate from the Hf$_2$CO$_2$ layer without any strain application. Notably, upon reaching a 2 $\%$ strain, a type-II transition becomes evident, with the blueP layer's contribution to the CBM becoming distinctly observable. At -5$\%$ strain, the VBM is predominantly contributed by the P-p states, while the CBM is primarily influenced by the Hf-d and O-p states. This finding strongly supports the transition from type-I to type-II behavior, as is also evident from the band-decomposed charge density plot. The changes in bandstructure and partial density of states of Zr$_2$CO$_2$/blueP under different biaxial strains are given in the Supporting Information Figure S4. Zr$_2$CO$_2$/blueP heterostructure also shows a similar pattern of band gap change while applying biaxial strain.

\begin{figure}[h!]
	\centering
	\includegraphics[width=\columnwidth]{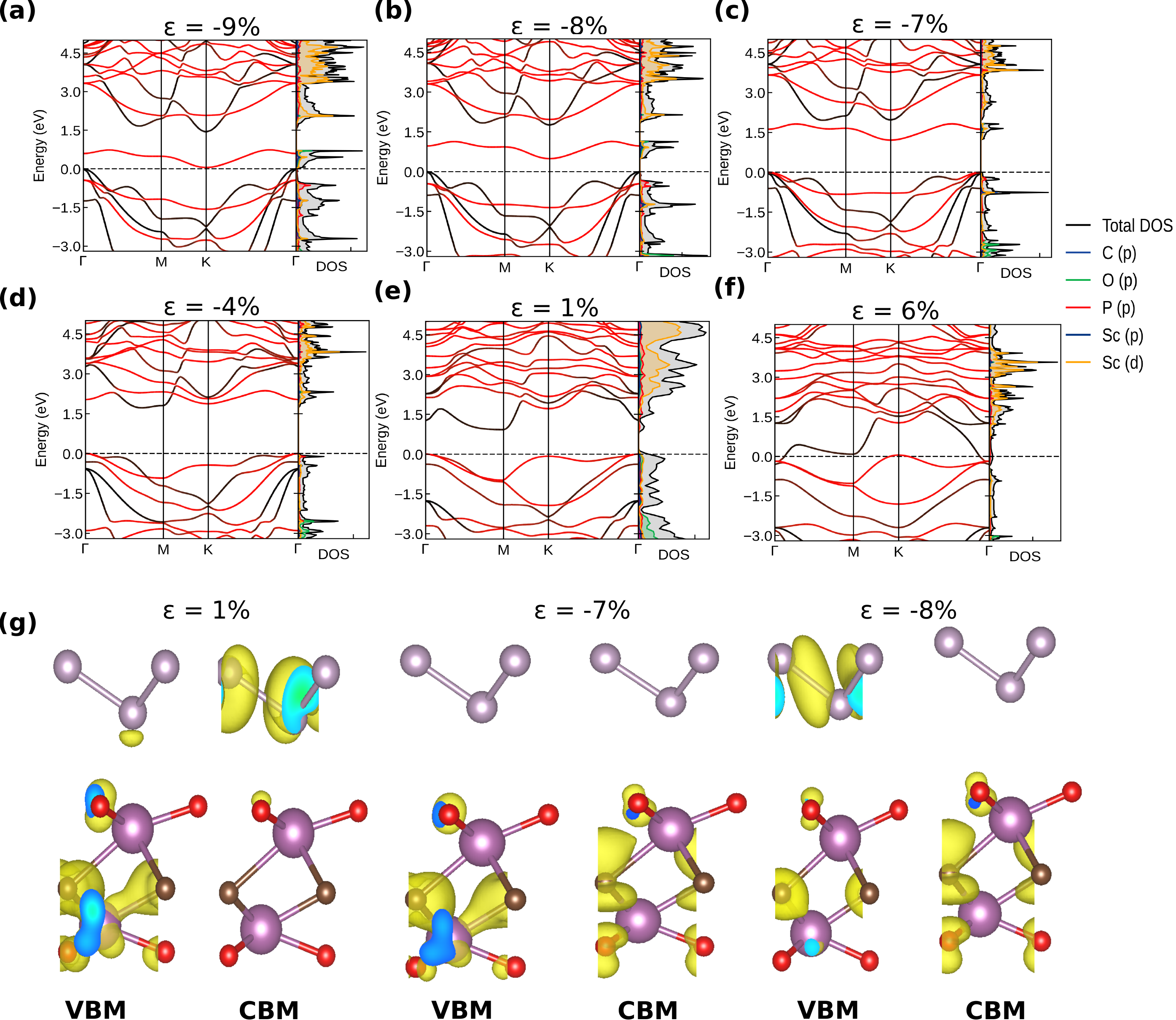}
	\caption{Band structures, PDOS  [(a)-(f)] and (g) Band decomposed charge density of Sc$_2$CO$_2$/blueP heterostructures under different percentages of biaxial strain}
	\label{fig9}
\end{figure}

The Sc$_2$CO$_2$/blueP heterostructure exhibits a different trend in band dispersion compared to the Hf$_2$CO$_2$/blueP and Zr$_2$CO$_2$/blueP heterostructures under biaxial strain, as shown in Figure  \ref{fig9}. From the band edge position under compressive strain, the position of CBM initially rises at the M point, reaching a maximum band gap of 1.7 eV at -4$\%$ strain. Beyond this point, the CBM shifts from M to K and begins to decrease, crossing the Fermi level at -9$\%$ strain. Concurrently, the VBM also moves towards the Fermi level with increasing strain. Under tensile strain, both the CBM and VBM shift towards the Fermi level as the strain increases. Initially, the VBM is at the $\Gamma$ point, but at 3$\%$ strain, it shifts to the K point and crosses the Fermi level, resulting in an S-M transition at 6$\%$. For better comparison, we have plotted the changes in the VBM and CBM under different biaxial strains, as shown in Figure S5 in the Supporting Information. The analysis of the partial density of states reveals significant insights into the behavior of the conduction and valence bands under strain. Throughout the application of tensile strain, we consistently observe the contribution of the p state of blueP to the CBM, while the VBM is primarily influenced by the Sc-d along with the O-p states. However, under compressive strain, a noticeable shift occurs. Up to -4$\%$ strain, the VBM is primarily contributed to by the Sc-d and the C-p states, while the CBM is influenced by the P-p states. Beyond -4$\%$, a gradual change is observed: the contribution of p state of blueP in the VBM and the Sc-d states in the CBM becomes more pronounced. At -5$\%$ strain, the VBM is entirely dominated by the p states of blueP, and the CBM is mainly influenced by Sc-d states. A notable transition from type-II to type-I behavior is observed at -7$\%$ strain, where both the VBM and CBM are dominated by Sc-d states. However, at -8$\%$ strain, the behavior reverts to type-II, with the VBM again dominated by P-p state and the CBM influenced by Sc-d states.\\

\begin{figure}[h!]
	\centering
	\includegraphics[width=\columnwidth]{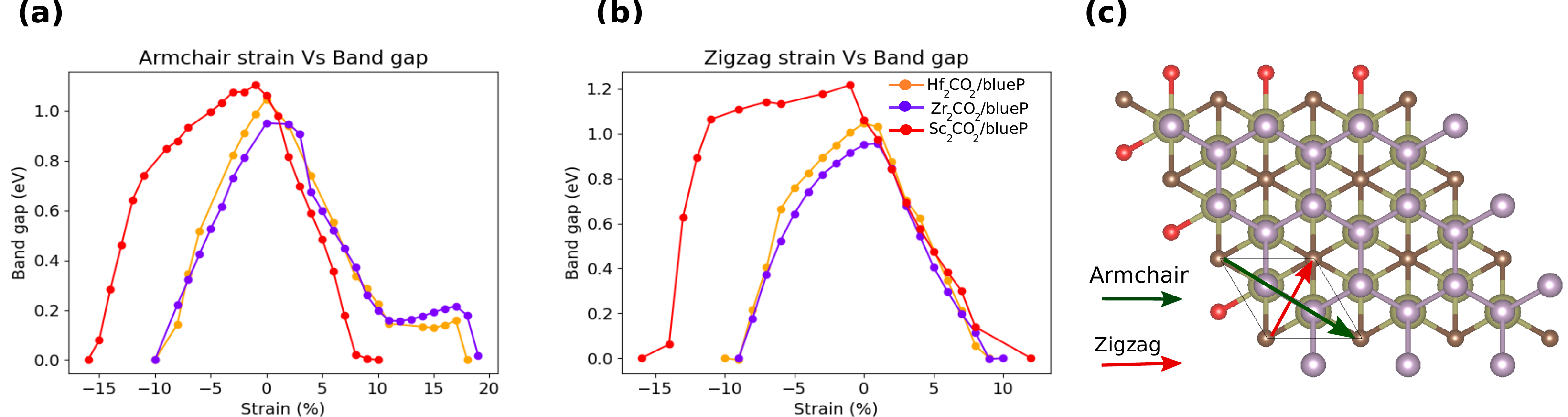}
	\caption{The change in bandgaps of MXene/blueP heterostructures under various percentages of applied uniaxial strain (a) armchair (b) zigzag, and (c) the schematic of uniaxial strain along the zigzag and armchair directions}
	\label{fig10}
\end{figure}

{\underline{\textbf{Uniaxial strain along zig-zag and armchair direction}}} \textbf{:} It has been already observed from the previous section that strain shows remarkable effects in modifying the electronic properties of M$_2$CO$_2$/blueP heterostructure. In this section, we will examine the effects of uniaxial strain applied along both the zigzag and armchair directions (see schematic Figure \ref{fig10}(c)) on the electronic properties of the three M$_2$CO$_2$/blueP heterostructures.  Applying strain along these directions provides a powerful means to tailor material properties, optimize device performance, and develop advanced technologies across various fields, including electronics and photonics. This approach enables the design of materials with specific characteristics to meet diverse application requirements \cite{he2013experimental}.

We found that the changes in the band gap for Hf$_2$CO$_2$/blueP and Zr$_2$CO$_2$/blueP under uniaxial strain along both the zigzag and armchair directions show similar trends (see Figure \ref{fig10}  When strain is applied along the zigzag direction, the band gap decreases for both tensile and compressive strain, with metallic transitions occurring when strain exceeds certain critical values. Although a similar S-M transition was observed for compressive strain along the armchair direction, no metallic transition was observed under tensile strain. Instead, the band gap reduces for the initial values of strain, but for larger values, there is a small increment in the band gap, which starts to decrease again reaching a minimal value of 0.016 eV for Zr$_2$CO$_2$/blueP at +19$\%$ strain.  Similarly, for Hf$_2$CO$_2$/blueP, the band gap reaches a minimal value of 0.003 eV at 18$\%$ strain. The band gap opens up again for larger strain for both the heterostructures. Sc$_2$CO$_2$/blueP shows a different behavior in a change in band gap under uniaxial strain than that of the other two heterostructures. For strain along armchair direction, the band gap was found to be decreasing and becoming metallic for tensile strain of 10 \%. Whereas for compressive strain, the band gap is found to be increasing first, then becomes constant for initial values of strain and after 4$\%$, the band gap started reducing and became metallic for 16$\%$ compressive strain. A similar trend was observed for strain along zig-zag directions too.

\begin{figure}[h!]
	\centering
	\includegraphics[width=\columnwidth]{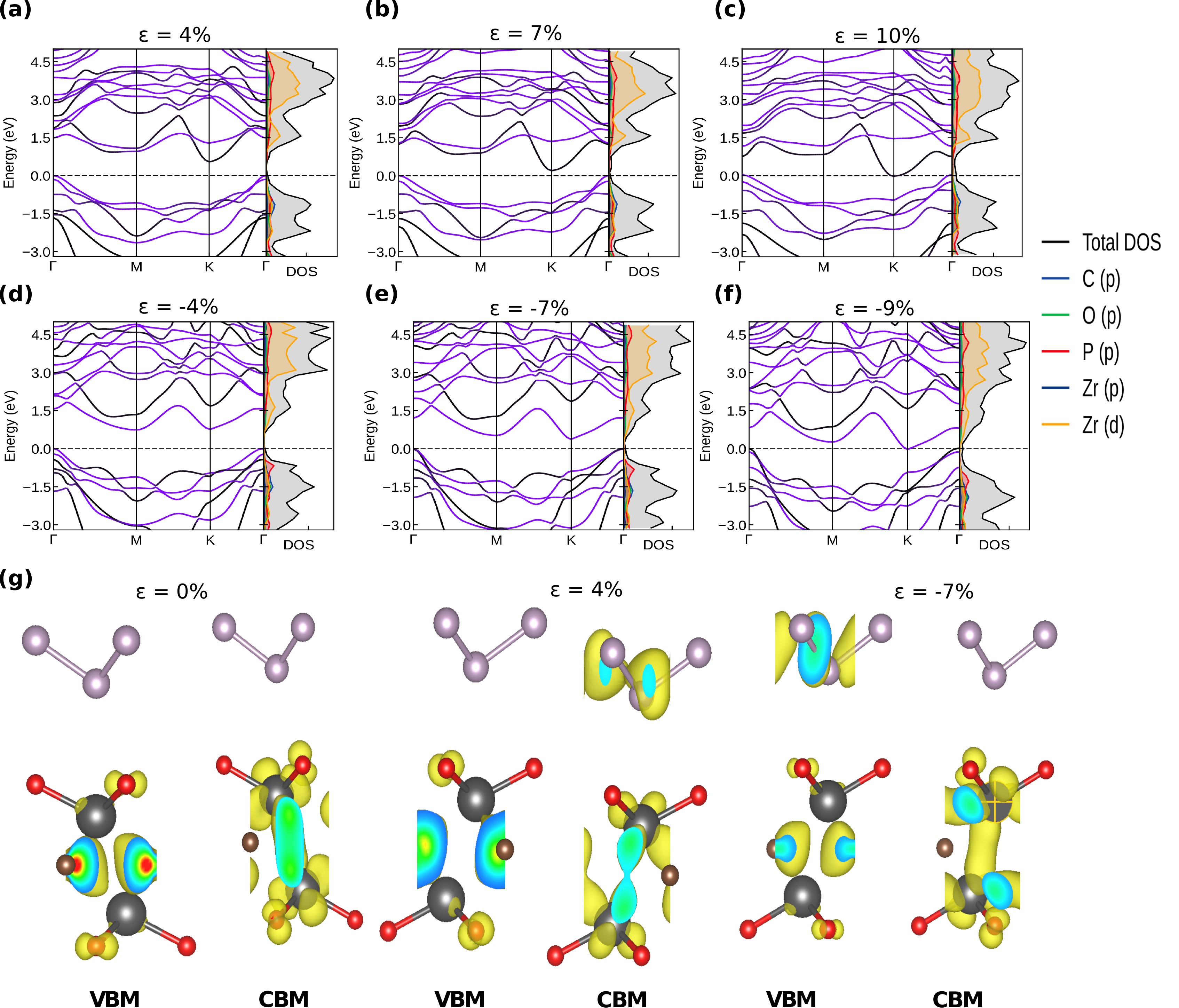}
	\caption{Band structures, PDOS  [(a)-(f)] and (g) Band decomposed charge density of Zr$_2$CO$_2$/blueP heterostructures under different percentages of uniaxial strain along zigzag direction}
	\label{fig11}
\end{figure}

\begin{figure}[h!]
	\centering
	\includegraphics[width=\columnwidth]{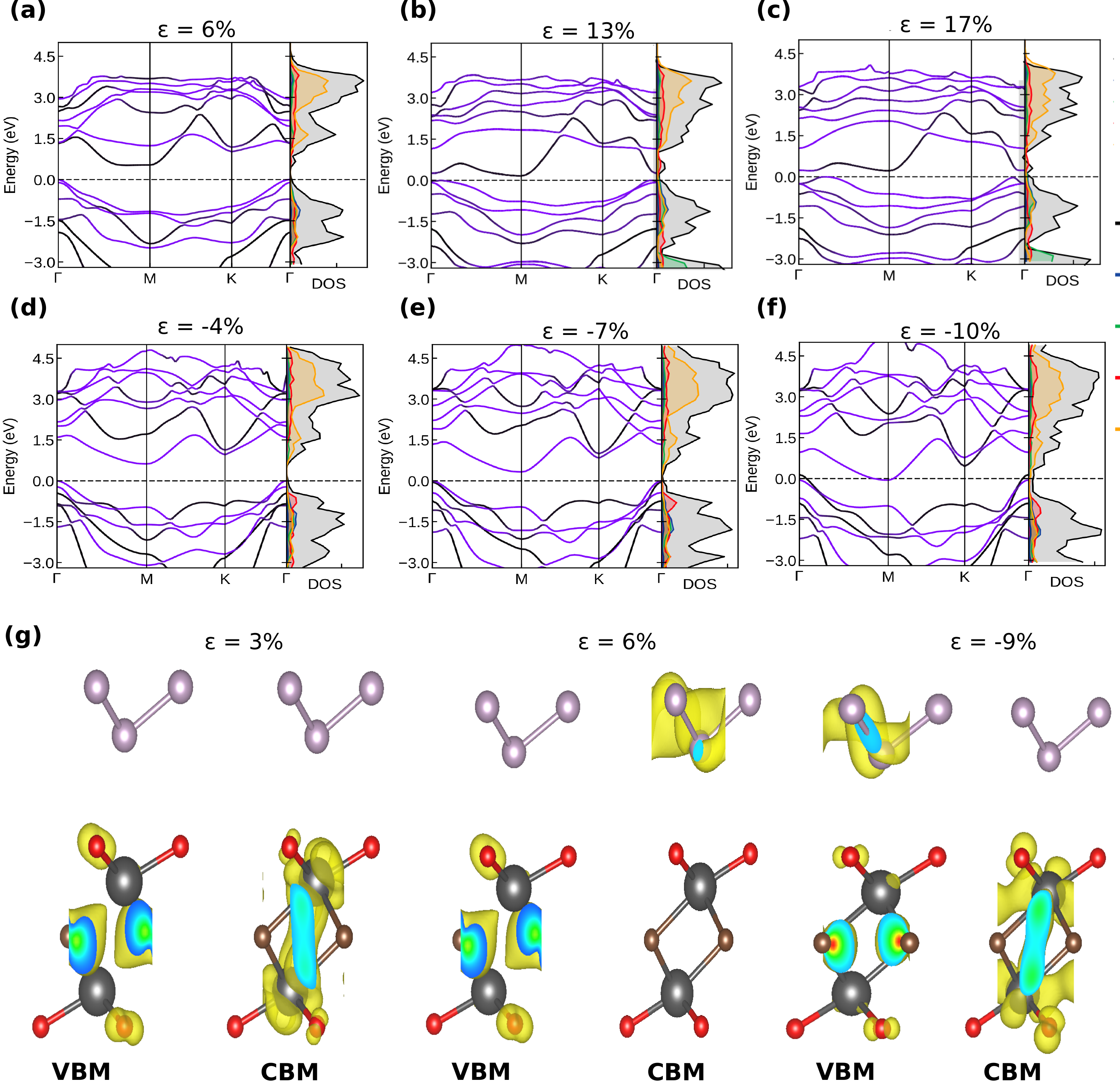}
	\caption{Band structures, PDOS  [(a)-(f)] and (g) Band decomposed charge density of Zr$_2$CO$_2$/blueP heterostructures under different percentages of uniaxial strain along armchair direction}
	\label{fig12}
\end{figure}

To observe the changes in band dispersions under uniaxial strain, we have plotted the band structures, PDOS, and band-decomposed charge density for Zr$_2$CO$_2$/blueP in the zigzag and armchair directions, as shown in Figures \ref{fig11} and \ref{fig12}, respectively. For strain applied in the zigzag direction, the CBM shifts closer to the Fermi level and crosses it, leading to a metallic transition for both compressive and tensile strain at the K point. From the band structure, it is visible that there is a transition from type-I to type-II for 7$\%$ of compressive strain and 4$\%$ of tensile strain. The band decomposed charge density plot for the Zr$_2$CO$_2$/blueP heterostructure under zigzag strain, as depicted in Figure  \ref{fig11} (g), offers valuable insights. Initially, at zero strain, both the VBM and CBM are primarily attributed to the Zr$_2$CO$_2$ layer, indicating a type-I heterostructure configuration. However, as the zigzag strain is applied, a transition to a type-II heterostructure becomes evident in both plots. The analysis of the partial density of states further elucidates the strain-dependent behavior. Up to 3$\%$ tensile strain, the VBM is chiefly influenced by the Zr-d states, along with contributions from the O-p and C-p states, while the CBM is predominantly governed by the Zr-d states. Beyond 3$\%$, the CBM increasingly reflects the influence of the p state of blueP while maintaining VBM dominance by the Zr-d states, where the type-I to type-II transition occurs. In the case of compressive strain, the CBM is primarily contributed to by the Zr-d states, while the VBM sees significant contributions from the C-p, Zr-d, and O-p states. Notably, at -7$\%$ strain, there's a marked increase in the involvement of the p state of blueP to the VBM, accompanied by a type-II transition.

The band structures for strain along the armchair direction are plotted in Figure  \ref{fig12}. There is a transition from type-I to type-II occurs for tensile strain as well as compressive strain. The CBM shifts towards the Fermi level at the M point  with an increase in both compressive and tensile strain, leading to a reduction in the band gap. The metallic transition happens when CBM (VBM) crosses the Fermi level at  M ($\Gamma$) point under compressive strain of 10\%. A shift in VBM position from $\Gamma$ point was observed when the tensile strain value was higher than 13\%. No metallic transition was observed within the tensile strain range applied here. The band-decomposed charge density for Zr$_2$CO$_2$/blueP, illustrated in Figure \ref{fig12}(g), indicates a type-II heterostructure under strain in the armchair direction. The PDOS analysis reveals significant findings regarding the behavior under armchair strain. Initially, at the onset of tensile strain, the VBM is primarily influenced by the C-p states and Zr-d state, while the CBM is largely contributed to by the Zr-d state. However, at 6$\%$ strain, there's a notable increase in the contribution of the p state of blueP to the CBM, and type-II transition occurs. Conversely, under compressive strain, at 9$\%$, there is a marked increase in the involvement of the p states of blueP in the VBM, also leading to a type-II transition

The change in the band structure of Hf$_2$CO$_2$/blueP under different percentages of strain in the zigzag and armchair directions is displayed in the Supporting Information (Figures S5 and S6). Applying 2$\%$ tensile strain and 7$\%$ compressive strain in the armchair direction induces a transition from type-I to type-II. For strain applied in the zigzag direction, the transition from type-I to type-II occurs at 1$\%$ tensile strain and 6$\%$ compressive strain. For the armchair direction, a S-M transition is observed only with compressive strain (10$\%$). The bandgap nearly closes at 18$\%$ tensile strain but starts to reopen at higher strains. Conversely, in the zigzag direction, the S-M transition occurs at 9$\%$ tensile strain and 10$\%$ compressive strain. The band decomposed charge density plot for different strain percentages along zigzag and armchair is shown in the Supporting Information Figure S6 and Figure S7 (g).

\begin{figure}[h!]
	\centering
	\includegraphics[width=\columnwidth]{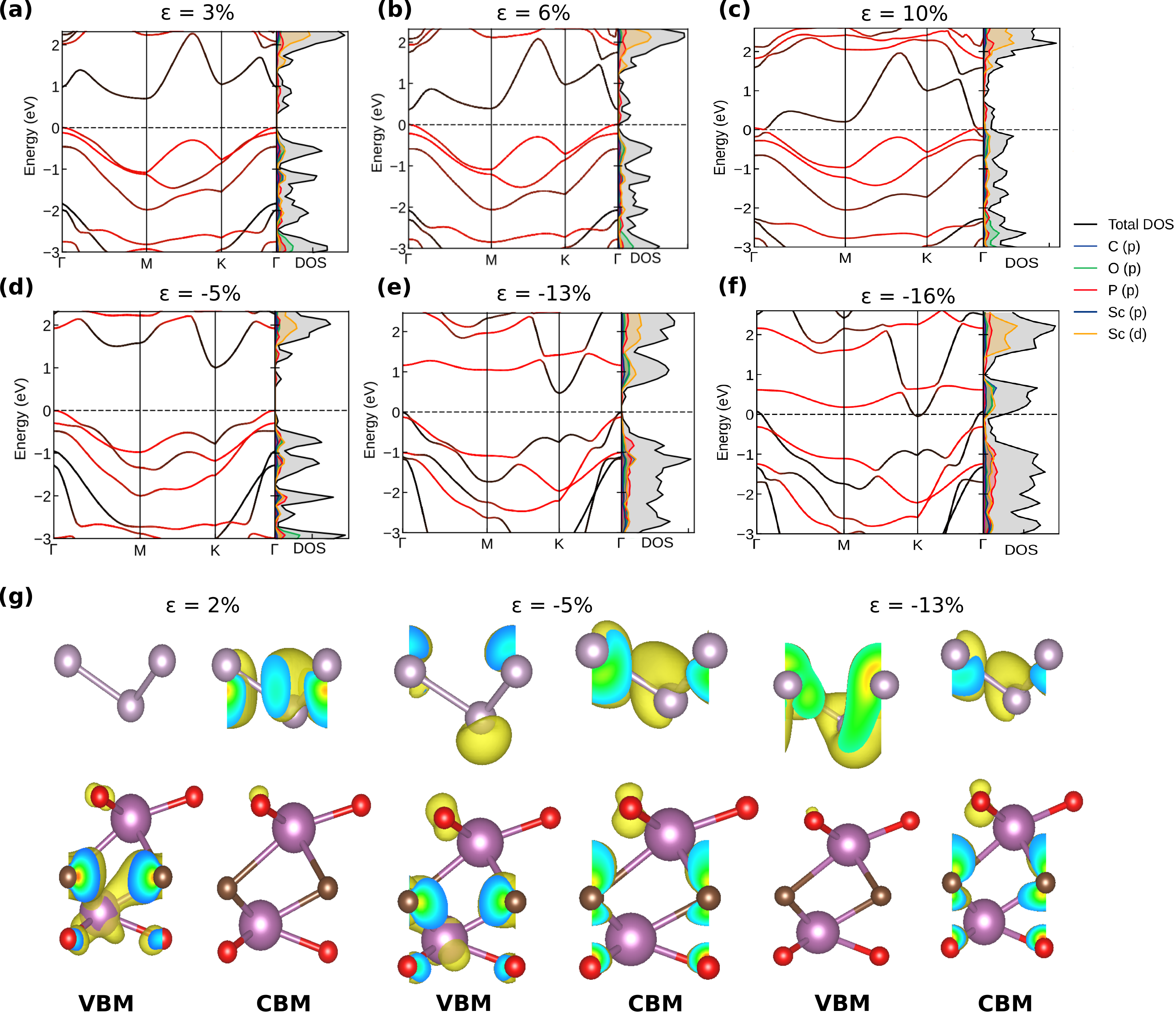}
	\caption{Band structures, PDOS  [(a)-(f)] and (g) Band decomposed charge density of Sc$_2$CO$_2$/blueP heterostructures under different percentages of uniaxial strain along armchair direction}
	\label{fig13}
\end{figure}
Figure \ref{fig13} shows the change in the band structure of the Sc$_2$CO$_2$/blueP heterostructure under different percentage of uniaxial strain in the armchair direction. For tensile strain, we found that at 6$\%$ of strain, there is a transition from indirect to direct band gap, and for the 10$\%$ the band gap closes when the CBM crosses the Fermi level at the $\Gamma$ point. In the case of compressive strain, the band is found to be constant for initial values of strain, and after -5$\%$ the conduction band started decreasing. CBM is mainly contributed by the p state of blueP and VBM is by Sc-d state.  At -13 $\%$ the contribution of p state of blueP has slightly increased in the VBM, but there is no type-II to type-I transition occurring. Then finally for -16$\%$, the S-M transitions occur when CBM and VBM cross the Fermi level at K and $\Gamma$ point. Similar changes in band structure were observed for strain applied along the zig-zag direction. Indirect to direct band gap transition at 6$\%$ strain and closing of band gap at for 12 $\%$ strain occurs under tensile strain. In the case of compressive strain, there is no major change occurs other than S-M transition at -16$\%$ strain. The change in band structure for strain in the zig-zag direction is given in the Supporting Information (Figure S8).

\begin{figure}[h!]
	\centering
	\includegraphics[width=\columnwidth]{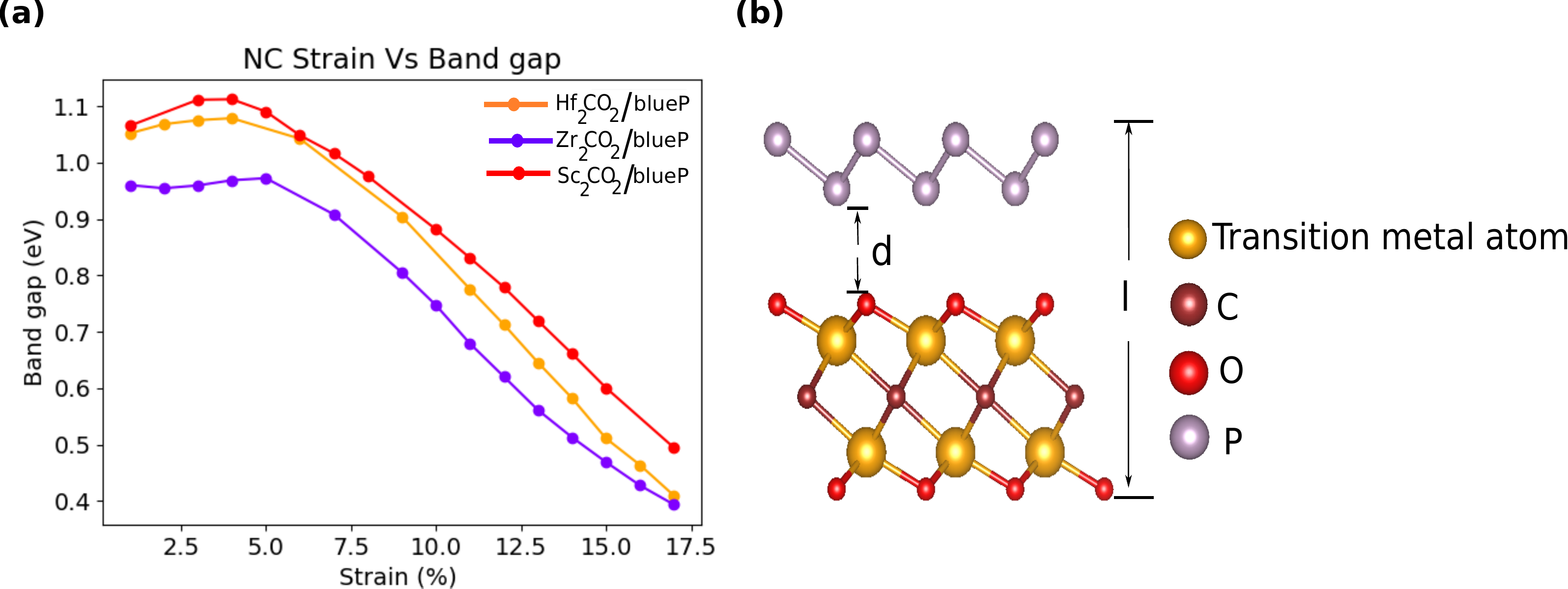}
	\caption{ (a) The change in bandgaps of MXene/blueP heterostructures under various percentages of applied NC strain and (b) Schematic of NC strain (Hf(Zr)$_2$CO$_2$/blueP heterostructure)}
	\label{fig14}
\end{figure}

\underline{\textbf{Normal compressive strain}}\textbf{:} Next, we have studied the impact of normal compressive (NC) strain  on the electronic properties of M$_2$CO$_2$/blueP heterostructures. The NC strain was determined as $\epsilon =({{l}_{0}}-l)/{{l}_{0}}$, where $l_0$ and $l$ are the un strained and strained layered distance (Figure  \ref{fig14} (b)), respectively. The interlayer distance is denoted by d. Figure \ref{fig14} (a) shows the calculated band gap as a function of NC strain. Initially, we observe a slight increment in the band gap for small strain percentages, which is followed by a smooth reduction as the strain increases. In contrast to other strains (biaxial, uniaxial), there is no S-M transition observed in all the considered heterostructures.

\begin{figure}[h!]
	\centering
	\includegraphics[width=\columnwidth]{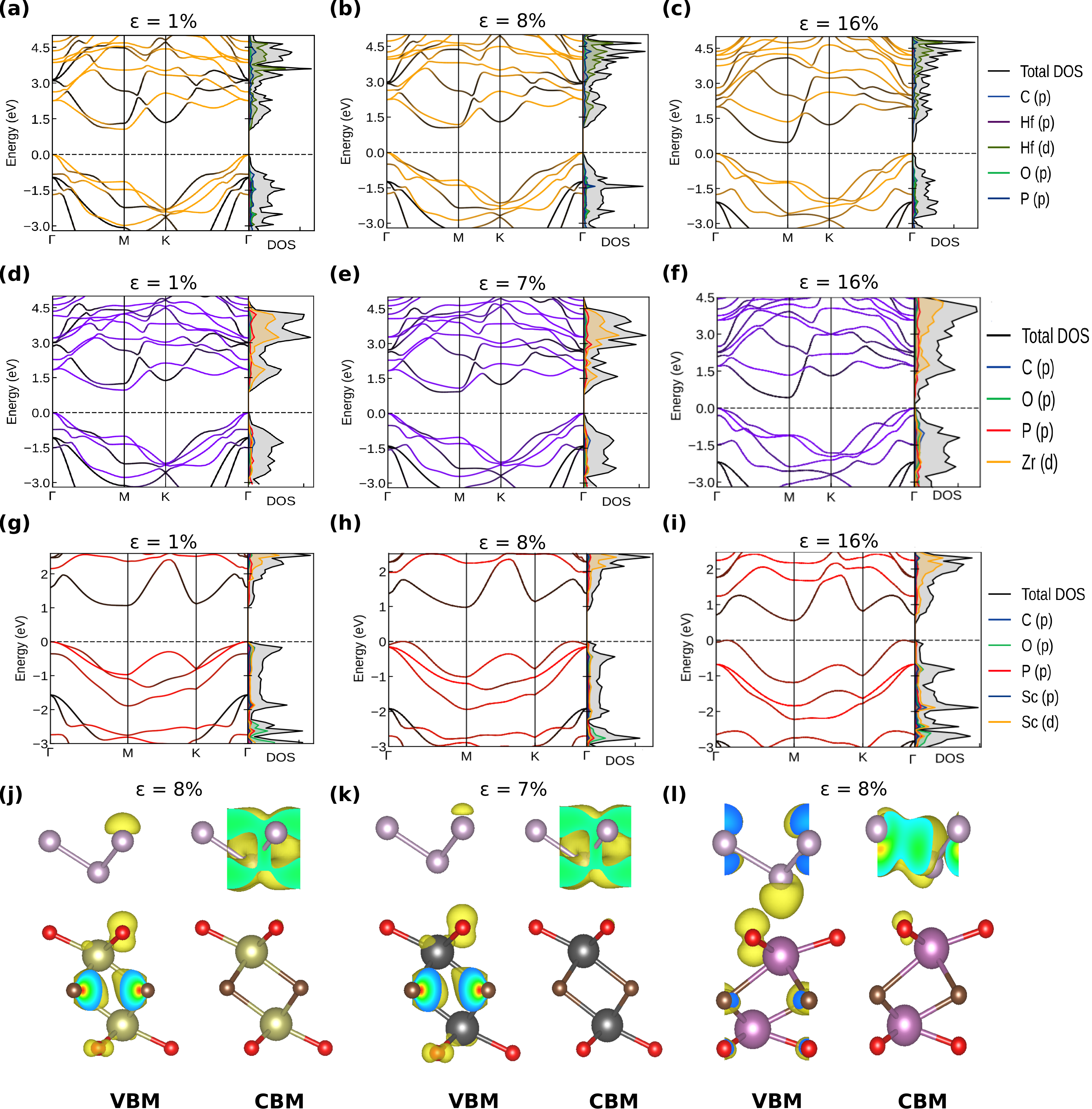}
	\caption{Band structures, PDOS of Hf$_2$CO$_2$/blueP [(a)-(c)], Zr$_2$CO$_2$/blueP [(d)-(f)], Sc$_2$CO$_2$/blueP [(g)-(i)] and band decomposed charge density of (j) Hf$_2$CO$_2$/blueP, (k) Zr$_2$CO$_2$/blueP, (l) Sc$_2$CO$_2$/blueP heterostructures\textcolor{red}{,} respectively under different percentage of NC strain}
	\label{fig15}
\end{figure}

The impact of NC strain on the electronic structure of each heterostructure was analyzed, as depicted in Figure \ref{fig15}. The band structure calculations show a consistent downward shift of the CBM across all considered systems, while the position of VBM remains unaffected in Hf$_2$CO$_2$/blueP and Zr$_2$CO$_2$/blueP under normal compressive strain, a notable shift occurs in Sc$_2$CO$_2$/blueP, where the VBM transitions from $\Gamma$ to in between $\Gamma$ and M points. Similar to the other strains, the normal compressive strain also causes the type-I to type-II conversion in Hf$_2$CO$_2$/blueP (8$\%$) and Zr$_2$CO$_2$/blueP (7$\%$) vdW heterostructures. The band decomposed charge density plot for M$_2$CO$_2$/blueP(M=Hf, Zr, Sc) is given in Figure  \ref{fig15} (j-l). In the Hf(Zr)$_2$CO$_2$/blueP heterostructure, the band gap decreases due to the CBM shifting closer to the Fermi level. Additionally, in the case of Sc$_2$CO$_2$/blueP heterostructure, under increased strain, shifts occur both in the VBM and CBM. For the initial strain conditions, the CBM of the Hf(Zr)$_2$CO$_2$/blueP heterostructure was primarily influenced by the Hf(Zr)-d states, while VBM saw significant contributions from the C-p and O-p states. As the NCS increased, the involvement of the p states of P in the CBM intensified. At 8(7)$\%$ strain, the dominance of the P-p state in the CBM was evident, leading to a transition from type-I to type-II behavior in the Hf(Zr)$_2$CO$_2$/blueP heterostructure. Conversely, in the Sc$_2$CO$_2$/blueP heterostructure, the CBM was largely governed by the p states of P, while the VBM saw notable contributions from the Sc-d, O(C)-p, and Sc-p states. With increasing strain, the participation of the p states of P in the VBM increased, although no dominance of these states was observed, and no type-II to type-I transition occurred.

\subsection{Thermoelectric properties}
Tuning of electronic properties of materials can influence their thermoelectric properties as these properties are closely related to the band structures. Utilizing band structures as previously computed, we have derived the electronic transport coefficients for all the heterostructures, encompassing the Seebeck coefficient (S), electrical conductivity ($\sigma/\tau$), electronic contribution to thermal conductivity ($\kappa_e/\tau$), power factor (S$^{2}\sigma/\tau$) and the electronic figure of merit, denoted by ZT$_e$, defined as:

\begin{equation}
    ZT_e = \frac{S^2\sigma T}{\kappa_e}
\end{equation}

This equation encapsulates the attributes of electron transport and serves as an upper limit for the ZT value\cite{mishra2020two}. According to the proportionality, ZT$_e$ can be maximized by increasing the numerator and decreasing the denominator. However, achieving this is complex due to the conflicting interdependencies among various coefficients. In high-temperature applications, thorough scrutiny of the influence of both power factor and thermal conductivity is essential, highlighting their paramount importance in such contexts\cite{wolf2019high}. Also, adjusting power factor (PF) has been demonstrated as a superior approach for enhancing efficiency and power generation compared to conventional methods of reducing lattice thermal conductivity, as it doesn't compromise thermo-mechanical stability\cite{liu2016importance}. We have explored the impact of doping charge carriers on these transport coefficients by adjusting the charge carrier concentrations. The negative and positive charge carrier concentrations correspond to electron (n-type) and hole (p-type) doping, respectively\cite{fecher2016half}.

 Seebeck coefficient, also referred to as thermopower, characterizes the electrical potential generated by a temperature gradient within a material. 
To achieve a high Seebeck coefficient, symmetry breaking is desired, often manifested as a sharp peak in the DOS. This can be accomplished through band convergence via strain application or resonant doping\cite{lim2021physical}. Electrical conductivity, denoted as $\sigma$, characterizes the medium's ability to conduct current. High electrical conductivity is a desirable characteristic of efficient thermoelectric materials. Although the CSTA  may oversimplify electrical transport phenomena, it provides a reasonable description of electrical conductivity, particularly under conditions such as doping of charge carriers or the application of external strain on the unit cell\cite{kumari2023strain}. 
 Electrical and thermal conductivities exhibit a proportional relationship as per Wiedemann-Franz law\cite{jones1985theoretical}, based on a simplistic free electron model. Considering the intricate interactions experienced by electrons due to atomic and electronic environments, there exists potential for optimizing the thermoelectric performance of materials. Herein, we systematically compute the electronic transport coefficients and their susceptibility to different strain conditions.

\begin{figure}[h!]
 	\centering
 	\includegraphics[width=\columnwidth]{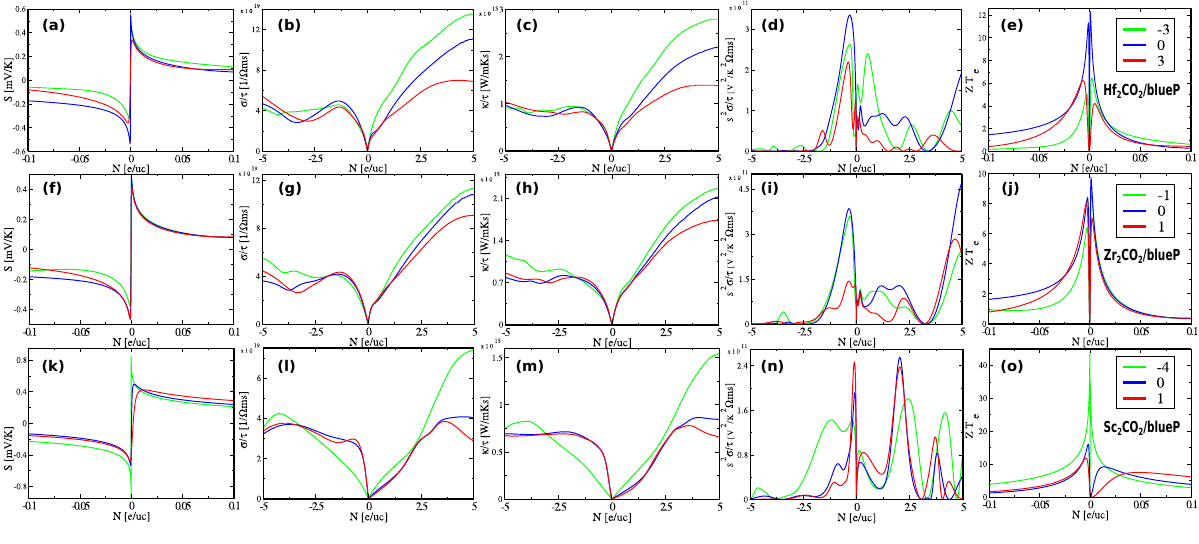}
 	\caption{Calculated Seebeck coefficient, Electrical conductivity, Thermal conductivity, Power Factor and electronic Figure of Merit of Hf$_2$CO$_2$/blueP [(a)-(e)], Zr$_2$CO$_2$/blueP [(f)-(j)] and Sc$_2$CO$_2$/blueP [(k)-(o)]  heterostructures respectively for different percentages of biaxial strain}
 	\label{fig_thermo1}
    \end{figure}

\underline{\textbf{Biaxial strain}} \textbf{:}
From the calculated seebeck coefficient, electrical conductivity, thermal conductivity, power factor and electronic figure of merit as shown in Figure  \ref{fig_thermo1}, it can be inferred that, the thermal and electrical conductivity profiles of both Hf$_2$CO$_2$/blueP and Zr$_2$CO$_2$/blueP  heterostructures exhibit notable similarities. In the case of p-type doping, positive strain induces a decrease in both conductivities, whereas negative strain results in an increase. In smaller doping concentration ranges, n-type doping largely has minimal effects on conductivities. However, as the doping concentration increases, significant changes in conductivity values become apparent. Comparing the Seebeck coefficients, Hf$_2$CO$_2$/blueP  heterostructures demonstrate greater variability compared to Zr$_2$CO$_2$/blueP  heterostructures, a trend also reflected in their respective ZT$_e$ values. ZT$_e$ reaches its maximum at zero strain for both  heterostructures. Conversely, for the PF, Zr$_2$CO$_2$/blueP  heterostructures demonstrate more variability than Hf$_2$CO$_2$/blueP when strain is introduced. Sc$_2$CO$_2$/blueP  heterostructure also displays analogous patterns in the effects of positive and zero strain on conductivities. However, negative strain profoundly impacts the conductivities in both doping domains. The Seebeck coefficient peaks at negative strain values, corresponding to a proportional increase in ZT$_e$ values.

\begin{figure}[htb!]
 	\centering
 	\includegraphics[width=\columnwidth]{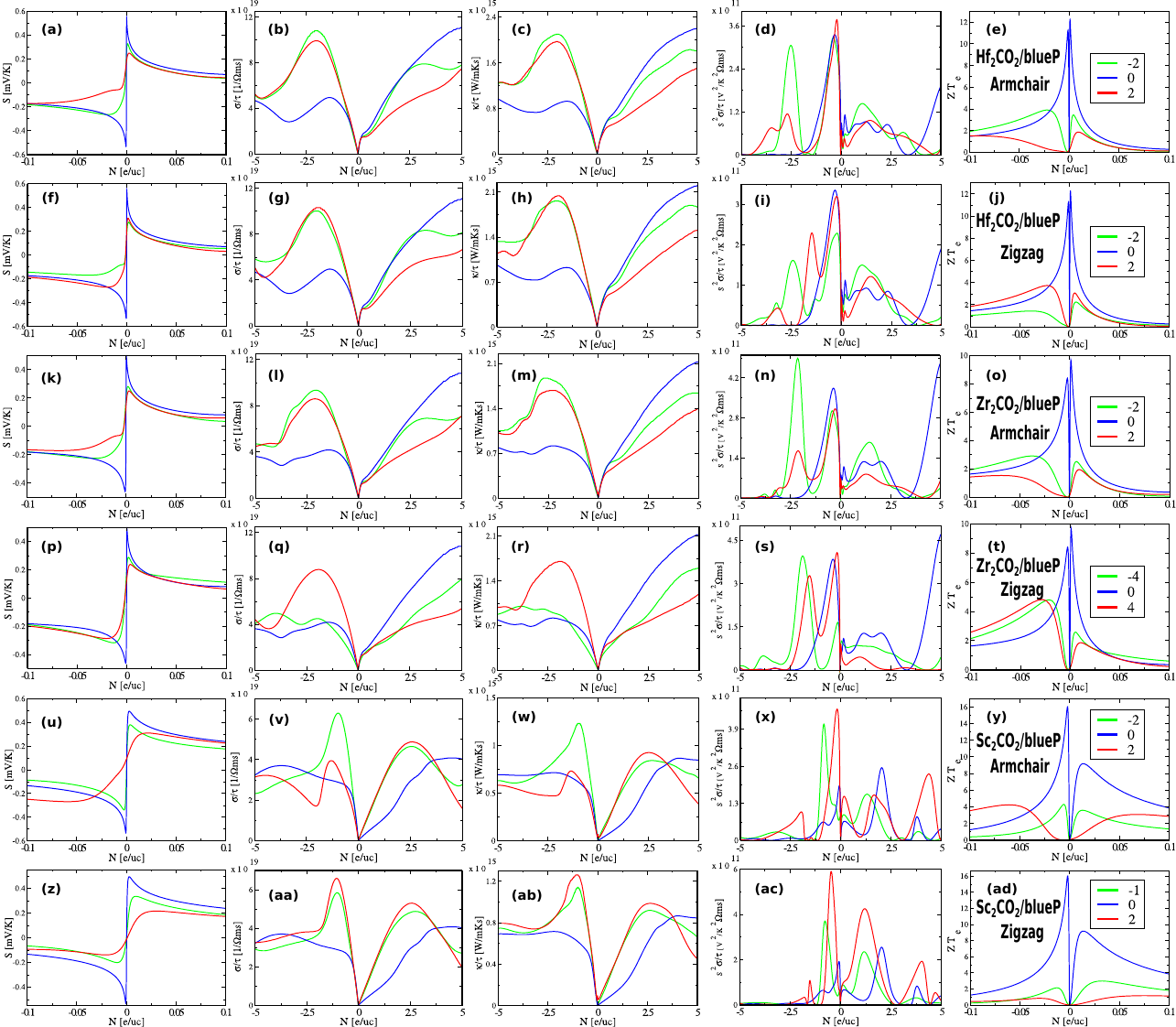}
 	\caption{Calculated Seebeck coefficient, Electrical conductivity, Thermal conductivity, Power Factor and electronic Figure of Merit of Hf$_2$CO$_2$/blueP [Armchair (a)-(e), Zigzag (f)-(j)], Zr$_2$CO$_2$/blueP [Armchair (k)-(o), Zigzag (p)-(t)] and Sc$_2$CO$_2$/blueP [Armchair (u)-(y), Zigzag (z)-(ad)] respectively for different percentages of uniaxial strain}
 	\label{fig_thermo2}
    \end{figure}

\underline{\textbf{Uniaxial strain along zig-zag and armchair direction}}\textbf{:}
Based on the calculated Seebeck coefficient, electrical conductivity, thermal conductivity, power factor, and electronic figure of merit presented in Figure  \ref{fig_thermo2}, the following observations can be inferred. For all the heterostructures, armchair and zigzag strain exert no positive influence on the Seebeck coefficient, a pattern also evident in the ZT$_e$ values. The maximum ZT$_e$ values are observed with zero strain across all combinations of heterostructures. Both strains lead to a decrease in the Seebeck coefficient, with positive strain causing a more pronounced reduction compared to negative strain. This same trend is observed in the ZT$_e$ values. Under negative armchair strain, both Hf$_2$CO$_2$/blueP and Zr$_2$CO$_2$/blueP heterostructure exhibit peaks around a charge carrier concentration of -2.5 in the conductivity plots, which correspondingly maximizes the power factor in the same regions. In the case of zigzag strain, while both strains decrease conductivities for both heterostructures in the p-type doping regions, they increase in the n-type regions. Significant discrepancies between positive and negative strain are observed for Zr$_2$CO$_2$/blueP heterostructure in the n-type regions, whereas the profiles remain relatively similar for Hf$_2$CO$_2$/blueP heterostructure. This observation serves as the evident reason for the power factor peaking at different regions for different strains in Zr$_2$CO$_2$/blueP heterostructures, while the power factor peaks around zero doping areas for Hf$_2$CO$_2$/blueP heterostructures.

Armchair strain in Sc$_2$CO$_2$/blueP heterostructure significantly influences conductivities within the n-type doping domain, leading to a peak in the power factor around these areas for both strains. Similarly, zigzag strain also exerts a considerable impact on conductivities. However, positive strain induces substantial effects in conductivities, consequently maximizing the power factor.

\begin{figure}[htb!]
 	\centering
 	\includegraphics[width=\columnwidth]{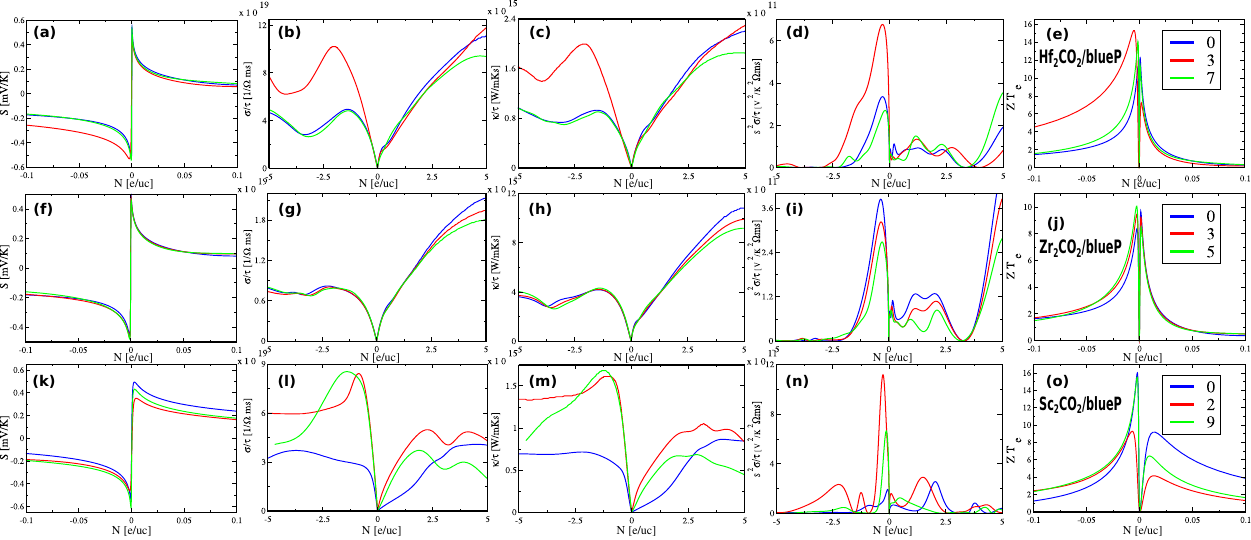}
 	\caption{Calculated Seebeck coefficient, Electrical conductivity, Thermal conductivity, Power Factor and electronic Figure of Merit of Hf$_2$CO$_2$/blueP [(a)-(e)], Zr$_2$CO$_2$/blueP [(f)-(j)], and Sc$_2$CO$_2$/blueP [(k)-(o)] heterostructures respectively for different percentages of normal compressive strain}
 	\label{fig_thermo3}
    \end{figure}

\underline{\textbf{Normal Compressive strain}}\textbf{:}
We have examined the impact of NC strain on the transport coefficients of M$_2$CO$_2$/blueP heterostructures as shown in Figure  \ref{fig_thermo3} .
In all the heterostructures, strain exhibits a positive effect on both power factor and ZT$_e$ values. Specifically in Hf$_2$CO$_2$/blueP heterostructure, at 3$\%$ strain, a significant disparity in the Seebeck coefficient and conductivity profiles emerges, leading to peaks in power factor and ZT$_e$ values in the n-type domains. In Zr$_2$CO$_2$/blueP heterostructure, minimal changes are observed in all the Onsager coefficient plots, resulting in minimal impact on both power factor and ZT$_e$. The power factor peaks for zero strain, whereas ZT$_e$ reaches its peak at 5$\%$ strain, where the Seebeck coefficient exhibits a slightly higher value compared to others. In Sc$_2$CO$_2$/blueP heterostructure, a significant increase in power factor is observed at 2$\%$ strain owing to a substantial rise in electrical conductivity. However, this leads to a decrease in ZT$_e$ due to a corresponding increase in thermal conductivity. Conversely, at 9$\%$ strain, the higher Seebeck coefficient and lower thermal conductivity result in an enlarged ZT$_e$ value.

The maximum PF, ZT$_e$, and band alignment type for all the heterostructures and their respective strain conditions are summarized in Table \ref{fig_thermo4}. 
The analysis indicates that negative biaxial strain has a profound effect on increasing the ZT$_e$ values specifically for the Zr$_2$CO$_2$/blueP heterostructure, surpassing all other heterostructures and strain conditions. Conversely, uniaxial strain, whether armchair or zigzag, did not lead to ZT$_e$ enhancement for any heterostructure. However, it did result in an increased power factor compared to the unstrained condition for all heterostructures except Hf$_2$CO$_2$/blueP in zigzag strain. Additionally, normal compressive strain notably doubled the power factor for Hf$_2$CO$_2$/blueP and increased it by over five times for Sc$_2$CO$_2$/blueP, marking the highest improvement among all heterostructures and different strain conditions. Our power factor values exhibit comparability and occasionally surpass those documented for existing thermoelectric materials\cite{das2023thermoelectric, ismail2022theoretical, bu2022dft}. Although ZT$_e$ values displayed some enhancement, Sc$_2$CO$_2$/blueP did not exhibit improvement in this aspect. In addition, we have provided band alignment diagrams to offer a clearer understanding of the band alignment types for both unstrained heterostructures and those under strain conditions where maximum PF and ZT$_e$ values are achieved, as given in the supplementary materials (see Figure S9).
\begin{table}[htb]
\begin{tabular}{| >{\centering\arraybackslash}m{1.1in} | >{\centering\arraybackslash}m{1.4in} | >{\centering\arraybackslash}m{1.2in} | >{\centering\arraybackslash}m{1.2in} |}
\hline
Heterostructure & Type of strain & Max PF \hspace{1cm} (x10$^{11}$ W/mK$^2$s) & Max ZT$_e$ \\ \hline
\multirow{4}{*}[1em]{\centering Hf$_2$CO$_2$/blueP} & Biaxial & 3.4 ($\epsilon$  = 0$\%$) (type-I) & 12.2 ($\epsilon$  = 0$\%$) (type-I) \\ \cline{2-4} 
 & Uniaxial (Armchair) & 3.8  ($\epsilon$  = 2$\%$)  (type-I) & 12.2 ($\epsilon$  = 0$\%$ )  (type-I)\\ \cline{2-4} 
 & Uniaxial (Zigzag) & 3.4  ($\epsilon$  = 0$\%$)  (type-I)& 12.2 ($\epsilon$  = 0$\%$)    (type-I)\\ \cline{2-4} 
 & Normal Compressive & \textbf{6.8}  ($\epsilon$  = 3$\%$)  (type-I) & \textbf{15.3} ($\epsilon$  = 3$\%$) (type-I)\\ \hline
\multirow{4}{*}[1em]{\centering Zr$_2$CO$_2$/blueP} & Biaxial & 3.8  ($\epsilon$  = 0$\%$) (type-I)& 9.8 ($\epsilon$  = 0$\%$) (type-I)\\ \cline{2-4} 
 & Uniaxial (Armchair) & \textbf{4.9}  ($\epsilon$  = -2$\%$)  (type-I) & 9.8 ($\epsilon$  = 0$\%$)  (type-I) \\ \cline{2-4} 
 & Uniaxial (Zigzag) & 4.1  ($\epsilon$  = 4$\%$)  (type-I) & 9.8  ($\epsilon$  = 0$\%$)   (type-I)\\ \cline{2-4} 
 & Normal Compressive & 3.8  ($\epsilon$  = 0$\%$)  (type-I) & \textbf{10.1} ($\epsilon$  = 5$\%$) (type-I) \\ \hline
\multirow{4}{*}[1em]{\centering Sc$_2$CO$_2$/blueP} & Biaxial & 2.6  ($\epsilon$  = 0$\%$) (type-II)& \textbf{43.2} ($\epsilon$  = -4$\%$)    (type-II)\\ \cline{2-4} 
 & Uniaxial (Armchair) & 4.6  ($\epsilon$  = 2$\%$)   (type-II)& 16.0 ($\epsilon$  = 0$\%$)    (type-II)\\ \cline{2-4} 
 & Uniaxial (Zigzag) & 5.9  ($\epsilon$  = 2$\%$)   (type-II) & 16.0 ($\epsilon$  = 0$\%$)    (type-II) \\ \cline{2-4} 
 & Normal Compressive & \textbf{11.2}    ($\epsilon$  = 2$\%$)   (type-II) & 16.0  ($\epsilon$  = 0$\%$)   (type-II) \\ \hline
\end{tabular}
\caption{Maximum PF, ZT$_e$, and band alignment type for all the heterostructures under different strain types and the corresponding strain values. The maximum obtained values of PF and ZT$_e$ under strain has been indicated in bold for each heterostructure}
\label{fig_thermo4}
\end{table}

\subsection{CONCLUSION}
We conducted a systematic study on the geometric structure, stability, electronic properties and modulation of electronic and thermoelectric properties of M$_2$CO$_2$ (M = Hf, Zr, Sc)/blueP heterostructures under biaxial and uniaxial strain.  Hf$_2$CO$_2$/blueP, Zr$_2$CO$_2$/blueP was observed to be type-I and  Sc$_2$CO$_2$/blueP a type-II heterostructure with indirect band gaps of 1.04 eV, 0.95 eV, and 1.06 eV, respectively.  Employing phonon dispersion and \textit{ab initio} molecular dynamics (AIMD) calculations, we confirmed the dynamical and thermal stability of all the heterostructures considered. Additionally, our determination of the work function, derived from electrostatic potential calculations, demonstrated that the MXene components possess lower work functions than blueP.

Further investigation on strain engineering of these heterostructures revealed interesting modulation of electronic and thermoelectric properties such as, semiconductor to metal (S-M) transition, type-I to type-II conversion and enhanced PF and ZT$_e$. Hf$_2$CO$_2$/blueP and Zr$_2$CO$_2$/blueP show remarkable similarities under the application of strain. Both heterostructures demonstrate a discernible reduction in band gap with increasing compressive and tensile biaxial strain, culminating in a S-M transition at -7$\%$ for both the heterostructures and at 18$\%$ and 14$\%$, for Zr$_2$CO$_2$/blueP and Hf$_2$CO$_2$/blueP,  respectively. In contrast, Sc$_2$CO$_2$/blueP shows an initial upward trend in band gap under low values of compressive strain, followed by a decrease, achieving an S-M transition at -10$\%$ strain. Under tensile strain, a sharp decrease in band gap is observed, with the heterostructure becoming metallic at 7$\%$ strain. We observed a type-I to type-II transition in both Hf$_2$CO$_2$/blueP and Zr$_2$CO$_2$/blueP under both tensile and compressive strain. Interestingly, we found a type-II to type-I transition in Sc$_2$CO$_2$/blueP at -7$\%$ strain, which shifts back to type-II at -8$\%$ strain. Applying uniaxial armchair strain to Hf(Zr)$_2$CO$_2$/blueP heterostructures did not close the band gap but induced a type I to type II transition, whereas zigzag strain induced both S-M and type I to type II transitions. For Sc$_2$CO$_2$/blueP, we observed an indirect-to-direct band gap transition under tensile strain for both armchair and zigzag orientations. We have observed S-M transition for both zigzag and arm chair under tensile and compressive strain. Notably, normal compressive (NC) strain did not induce a S-M transition but led to a gradual decrease in the band gap for large strain values. Similarly, as with other strains, a type I to type II transition was observed for Hf(Zr)$_2$CO$_2$/blueP heterostructures under NC strain.

In an extended study of strain engineering on thermoelectric properties, we observed a positive correlation between strain and power factor in the majority of strain conditions, whereby the electronic figure of merit is predominantly negatively affected under strain, with notable exceptions observed for biaxial strain, where its influence is substantial. Sc$_2$CO$_2$/blueP showed the highest ZT$_e$ value of 43.2 at -4\% biaxial strain  among all the heterostructures. The maximum power factor (x10$^{11}$ W/mK$^2$s) obtained for Hf$_2$CO$_2$/blueP, Zr$_2$CO$_2$/blueP and Sc$_2$CO$_2$/blueP are 6.8 ( at 3\% NC strain), 4.9 (at -2\% uniaxial armchair strain) and 11.2 (at 2\% NC strain), respectively. We contend that the fine-tuning of the power factor holds paramount significance in high-temperature applications, and our discoveries are poised to be particularly valuable in this domain. These results highlight the potential of M$_2$CO$_2$/blueP heterostructures for electronic, optoelectronic and thermoelectric applications.

\subsection*{CRediT authorship contribution statement}
\textbf{Sarga P K (First Author)}: Conceptualization, Data Curation, Methodology, Investigation, Formal Analysis, Writing - Original Draft, Review \& Editing. \textbf{Karthik H J}: Investigation, Methodology, Data curation, Formal Analysis, Writing \textcolor{red}{ -} Review \& Editing. \textbf{Swastibrata Bhattacharyya (Corresponding Author)}: Conceptualization, Methodology, Validation, Formal Analysis,  Resources, Supervision,  Writing - Review \& Editing.

\begin{suppinfo}

\begin{itemize}
\item{Lattice parameter and bond lengths of all considered monolayers; Formation energy and Bader charge distribution in M$_2$CO$_2$/blueP heterostruture; Top and side views of the different stacking configurations of the M$_2$CO$_2$/blueP heterostructure; Elastic constants (C$_{11}$, C$_{12}$, C$_{66}$), and Young’s modulus(Y) of monolayers
and heterostructures; Band structure, PDOS and Band decomposed charge density of Zr$_2$CO$_2$/blueP (Hf(Sc)$_2$CO$_2$/blueP) heterostructure under different percentages of biaxial (Zigzag and armchair) strain}; Band alignment of M$_2$CO$_2$/blueP heterostructure for unstrained as well as strained values of maximum PF and ZT.
\end{itemize}

\end{suppinfo}

\begin{acknowledgement}
 SB acknowledges the financial support from the Science and Engineering Research Board (SERB), Government of India (grant numbers SRG/2020/000562, CRG/2020/000434 and CRG/2023/003688), and BITS Pilani K. K. Birla Goa Campus, India (grant numbers GOA/ACG/2019-20/NOV/08 and C1/23/211). SPK acknowledges the financial support from SERB, Government of India (grant number SRG/2020/000562). All authors acknowledge the High-Performance Computing Facility (HPC) at BITS Pilani K. K. Birla Goa Campus.
\end{acknowledgement}

\subsection*{Data availability}
Data will be made available on request.


\end{document}